\documentclass[11pt]{article}
\usepackage{fullpage,mathptm,times,amsmath,mathdots,amssymb,epsf,graphicx}
\advance\parskip 4pt
\usepackage{epstopdf}

\newtheorem{theorem}{Theorem}[section]
\newtheorem{definition}[theorem]{Definition}
\newtheorem{lemma}[theorem]{Lemma}
\newtheorem{proposition}[theorem]{Proposition}

\def\proof{\par\kern-\medskipamount\noindent\textbf{Proof.}~~}

\def\remark{\par\kern\medskipamount\noindent\textbf{Remark~\stepcounter{theorem}\arabic{section}.\arabic{theorem}}~}

\catcode`\@ 11
\@addtoreset{equation}{section}
\long\def\@makecaption#1#2{\vskip\abovecaptionskip
  \sbox\@tempboxa{\small #1: #2}%
  \ifdim \wd\@tempboxa >\hsize \small #1: #2\par
  \else \global \@minipagefalse \hb@xt@\hsize{\hfil\box\@tempboxa\hfil}\fi
  \vskip\belowcaptionskip}
\newenvironment{example}{\refstepcounter{theorem}
  \par\kern\medskipamount\noindent\textbf{Example~\thetheorem}~}{\par\smallskip}
\newenvironment{property}{\refstepcounter{theorem}
  \par\kern\medskipamount\noindent\textbf{Property~\thetheorem}~}{\par\smallskip}
\newenvironment{condition}{\refstepcounter{theorem}
  \par\kern\medskipamount\noindent\textbf{Condition~\thetheorem}~}{\par\smallskip}

\numberwithin{figure}{section}
\catcode`\@ 12

\let\==\bar
\let\@=\mathbf
\newcommand{\partialderiv}[3][]{\frac{\partial^{#1}#2}{\partial {#3}^{#1}}}
\def\circ{\ifmmode\mathchar"220E\else$\mathchar"220E$\fi}
\def\Wr{\mathop{\mathrm{Wr}}\nolimits}
\def\diag{\mathop{\rm diag}\nolimits}
\def\Real{{\mathbb{R}}}
\def\sech{\mathop{\rm sech}\nolimits}
\def\half{{\textstyle\frac12}}
\def\rank{\mathop{\rm rank}\nolimits}
\let\next=\phi\global\let\phi=\varphi\global\let\varphi=\next
\def\mapto#1#2{\mathop{\longrightarrow}\limits^{#1\to#2}} 
\renewcommand\labelitemi{\ifmmode\circ\else$\circ$\fi}
\allowdisplaybreaks

\begin{document}
\title{\bf Classification of the line-soliton solutions of KPII}
\author{Sarbarish Chakravarty$^1$ and Yuji Kodama$^2$ \\[1ex]
\small\it\
$^1$Department of Mathematics, University of Colorado, Colorado Springs, CO 80933 \\
\small\it\
$^2$ Department of Mathematics, Ohio State University, Columbus, OH 43210}
\date{}
\maketitle
\begin{abstract}
In the previous papers (notably, Y. Kodama, J. Phys. A 37, 11169-11190 (2004),
and G. Biondini and S. Chakravarty, J. Math. Phys. 47 033514 (2006)),
we found a large variety of line-soliton solutions of the
Kadomtsev-Petviashvili II (KPII) equation. The line-soliton solutions
are solitary waves which decay exponentially in $(x,y)$-plane
except along certain rays.  In this paper,
we show that those solutions are classified by asymptotic information
of the solution as  $|y| \to \infty$. Our study then
unravels some interesting relations between the line-soliton classification
scheme and classical results in the theory of permutations. 
\end{abstract}

\tableofcontents
\section{The KPII equation and its line-soliton solutions}
\label{s:introduction}

The Kadomtsev-Petviashvili (KP) equation
\begin{equation}
\partialderiv{}x\left(-4\partialderiv ut
  +\partialderiv[3]ux +6u\partialderiv ux \right)
            + 3\sigma^2 \partialderiv[2]uy =0\,,
\label{e:KP}
\end{equation}
where $u=u(x,y,t)$ and $\sigma^2=\pm1$,
describes the evolution of small-amplitude, quasi two-dimensional 
solitary waves in a weakly dispersive medium \cite{SovPhysDoklady15p539}.
The case $\sigma^2=-1$ corresponding to positive dispersion is known as 
the KPI equation, whereas the negative dispersion ($\sigma^2=1$) case is
referred to as the KPII equation. The KP equation arises in many physical 
applications including water waves and plasmas 
(see e.g. ~\cite{InfeldRowlands} for a review).
It is a completely integrable system with remarkably rich
mathematical structure which is well-documented in several monographs
~\cite{AblowitzClarkson,Hirota,MatveevSalle,MiwaJimboDate,NMPZ1984}.
Particularly, it has been 
known that the solutions of the KP equation can be expressed in terms of the $\tau$-function
\cite{Hirota, Sato},
\begin{equation}
u(x,y,t)= 2\partialderiv[2]{ }x\log\tau(x,y,t)\,.
\label{e:u}
\end{equation}
In this paper, we consider a class of solutions whose $\tau$-function is given by the Wronskian 
determinant~\cite{Sato,PLA95p1}
\begin{equation}
\tau(x,y,t)= \Wr(f_1,\dots,f_N)= 
  \begin{pmatrix}
     f_1 & f_2 & \cdots & f_N \\ 
     f_1' & f_2' &\cdots & f_N' \\
     \vdots & \vdots & & \vdots \\
     f_N^{(N-1)} & f_2^{(N-1)} & \cdots &f_N^{(N-1)}
  \end{pmatrix}\,.
\label{e:tau}
\end{equation}
with $f^{(i)}= \partial^i \!f/\partial x^i$,
and where the functions $\{f_n\}_{n=1}^N$
is a set of linearly independent solutions of the linear system
\begin{equation}\label{e:fpde}
\partialderiv fy= \partialderiv[2]fx\,,
\qquad
\partialderiv ft= \partialderiv[3]fx\,.
\end{equation}
In particular, we investigate the {\em line-soliton}
solutions of the KPII equation, which are real, non-singular
solutions localized along certain directions in the $(x,y)$-plane,
and decay exponentially everywhere else.
For example, a one-soliton solution is obtained by choosing $N=1$ 
in Eq.~ \eqref{e:tau} above, and $\tau (x,y,t) = f(x,y,t)= e^{\theta_1}+e^{\theta_2}$,
where
\begin{equation}
\theta_m(x,y,t) = k_mx +k_m^2y+k_m^3t+\theta_{m,0}
\label{e:theta}
\end{equation}
with $\theta_{m,0}, \, k_m$ for $m=1,2$ are constants, and $k_1 < k_2$.
The above choices yield the traveling-wave solution
\begin{equation}
u(x,y,t)= \half(k_2-k_1)^2\sech^2\half(\theta_2-\theta_1)= \Phi(\@k\cdot\@r+\omega t) \,,
\label{e:onesoliton}
\end{equation}
where $\@r=(x,y)$.
The wave vector $\@k:=(l_x,l_y) =(k_1-k_2,k_1^2-k_2^2)$ and the frequency $\omega$
satisfy the dispersion relation,
\begin{equation}
-4\omega l_x+l_x^4+3l_y^2=0\,. 
\label{e:dispersionrelation}
\end{equation}
The solitary wave given by Eq.~ \eqref{e:onesoliton} is localized 
in the $(x,y)$-plane along the line $L: \theta_1=\theta_2$
whose normal has the slope $c=l_y/l_x = k_1+k_2$. 
The one-soliton solution is characterized by two physical parameters, namely,
the \textit{soliton amplitude}~$a=k_2-k_1$ and the \textit{soliton direction}
$c=k_1+k_2$. The soliton direction can be also expressed as $c=\tan \alpha$,
where $\alpha$ is the angle, measured counterclockwise,
between the line $L$ and the positive $y$-axis.
Conversely, any given choice of amplitude ($a>0$) and direction of the soliton
gives the phase parameters $k_1$ and $k_2$ uniquely as
$k_1=\half (c-a)$ and $k_2=\half (c+a)$. 
Note that when $c=0$ (equivalently, $k_1 = -k_2$), the solution
in Eq.~\eqref{e:onesoliton} becomes $y$-independent and reduces
to the one-soliton solution of the Korteweg-de~Vries (KdV) equation.

\kern-\medskipamount
\paragraph{General line-soliton solutions.}
Like KdV, the KPII equation also admits multi-soliton solutions
which can also be constructed via the Wronskian formulation
of Eq.~\eqref{e:tau} by choosing~$M$ phases
$\{\theta_m\}_{m=1}^M$ defined as in Eq.~\eqref{e:theta}
with distinct real \textit{phase parameters} $k_1 < k_2 < \ldots < k_M$
and then defining the functions 
\begin{equation}
f_n(x,y,t)= \sum_{m=1}^{M} a_{nm}\,e^{\theta_m}\,, \quad
n = 1,2, \ldots, N\,,
\label{e:f}
\end{equation}
which give finite dimensional solutions of Eqs.~\eqref{e:fpde}.
The constant coefficients $a_{nm}$ define the $N \times M$
\textit{coefficient matrix} $A:= (a_{nm})$, all of whose $N \times N$ 
minors must be non-negative to ensure that the $\tau$-function $\tau(x,y,t)$
has no zeros in the $(x,y)$-plane for all $t$,
so that the corresponding KPII solution $u(x,y,t)$ resulting from Eq.~\eqref{e:u} is
non-singular.

\begin{figure}[t!]
\centering
\raisebox{0.85in}{(a)}\includegraphics[scale=0.55]{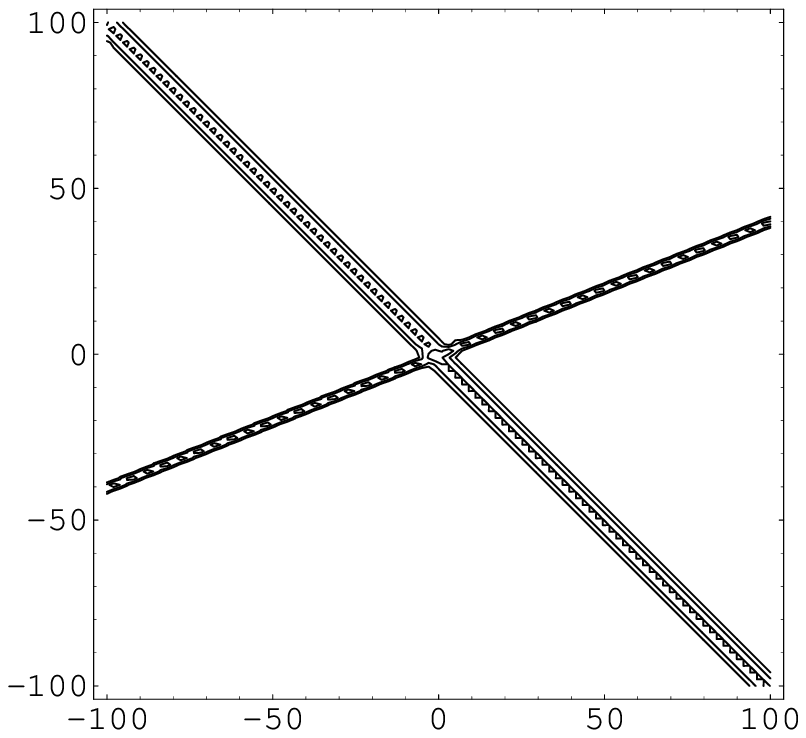} \hskip 0.3cm 
\raisebox{0.85in}{(b)}\raisebox{-0.1cm}{\includegraphics[scale=0.55]{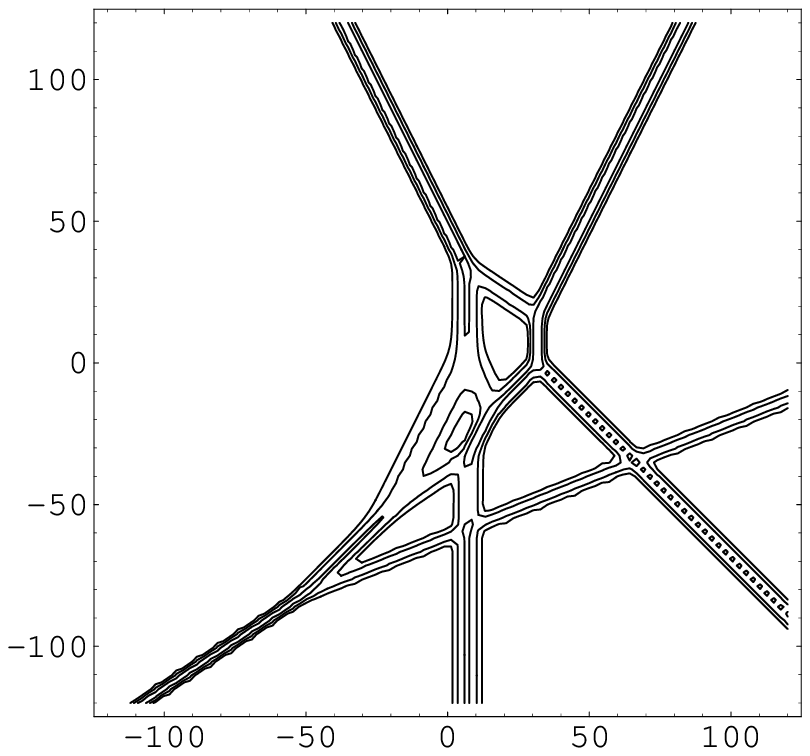}}\hskip 0.5cm
\raisebox{0.85in}{(c)}\includegraphics[scale=0.55]{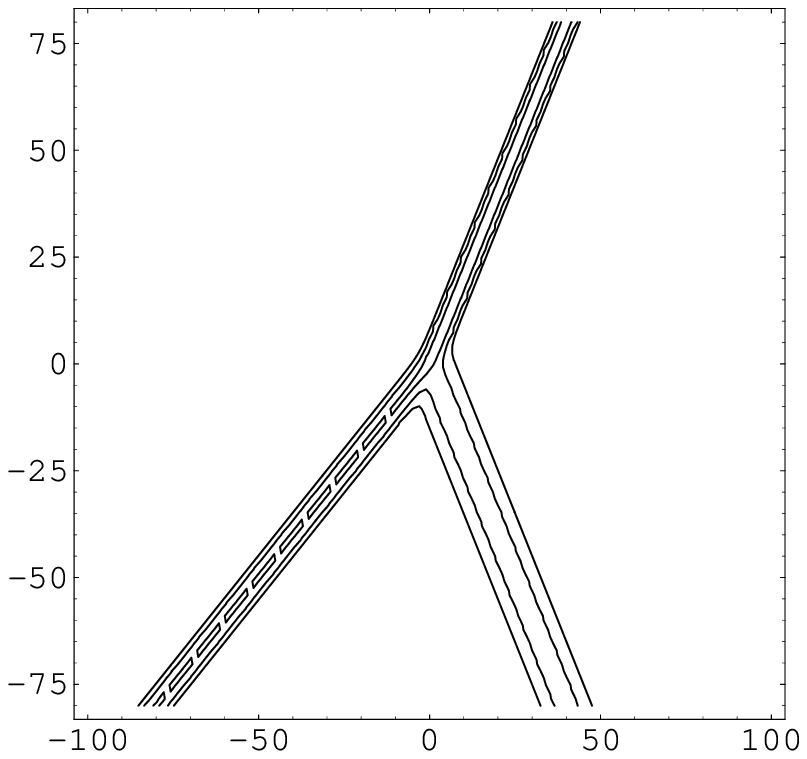} 
\caption{Line-soliton solutions of the KPII equation illustrating
different interaction patterns:
(a)~a 2-soliton solution,
(b)~a partially resonant $(3,3)$-soliton,
(c)~a Miles resonance (Y-junction).
Here and in all following figures, the horizontal and vertical axes are respectively,
$x$~and~$y$, and the graphs show contour lines of the solution
$u(x,y,t)=2\partial_x^2\,\log\tau(x,y,t)$ for fixed~$t$.}
\label{f:kpfig}
\end{figure}

However, the multi-soliton solution space of the KPII equation
turns out to be much richer than that of the (1+1)-dimensional KdV
equation due to the dependence of the KPII solutions on
the additional spatial variable $y$.
 Asymptotically as $y \rightarrow \pm\infty$, 
there exist certain (non-decaying) directions which are invariant in
$t$, and along which the solution 
has the form of a plane wave similar to the one-soliton solution in 
Eq.~\eqref{e:onesoliton}. These asymptotic solitary wave structures, 
referred to as {\em asymptotic} line-solitons in Ref.~\cite{BC},
have varying amplitudes and directions depending on $M$, $N$ and the
the values of the phase parameters $k_1,\dots,k_M$. More significantly,
the number $N_-$ of asymptotic line-solitons
as $y \rightarrow -\infty$ is in general different from the
number $N_+$ of the asymptotic line-solitons as $y \rightarrow \infty$,
with $N_- = M-N$ and $N_+=N$. Such multi-soliton configurations derived
from Eq.~\eqref{e:f} are called $(N_-,N_+)$-soliton solutions of KPII \cite{BC, jphysa36p10519}.
For example, Figure~\ref{f:kpfig}(c) exhibits a $(2,1)$-soliton solution,
also known as the Miles resonance solution~\cite{JFM1977v79p171}. 
At the interaction vertex or Y-junction, the three interacting line-solitons 
with wave numbers~$\@k_a$ and frequencies~$\omega_a$ ($a=1,2,3$)
satisfy the fundamental three-wave resonance condition
\begin{equation}
\@k_1 + \@k_2 = \@k_3\,,\qquad
\omega_1 + \omega_2 = \omega_3\,
\label{e:resonance}
\end{equation}
The $(N_-,N_+)$-soliton solutions exhibit a variety of time-dependent spatial 
interaction patterns including the formation of intermediate line-solitons 
in the $(x,y)$-plane ~\cite{jphysa36p10519,Kodama,medina}.
In contrast to these nontrivial interactions exhibited
by the KPII solitons, the KdV multi-soliton solutions experience only
a phase shift after collision.
\kern-\medskipamount
\paragraph{$N$-soliton solutions.}
When $N_-=N_+=N$ (i.e., when $M=2N$) the corresponding solutions consist of 
the same number of asymptotic line-solitons as $y \rightarrow \pm \infty$.
If in addition, the direction and amplitude of each of the $N$ line 
solitons as $y\to-\infty$ are {\em pairwise} equal to each of the 
$N$ line-solitons as $y\to\infty$, then the corresponding solutions
are simply referred to as the $N$-soliton solutions of the KPII equation.
It will be evident from the discussions in the following sections
that each $N$-soliton solution can be regarded as a configuration of
$N$ interacting asymptotic line-solitons where the amplitude and direction 
of the $n$-th line-soliton are given by 
\begin{equation}
a_n= k_{j_n}-k_{i_n},\quad c_n= k_{i_n}+k_{j_n},\qquad n=1,\dots,N\,.
\label{e:solitonparameters}
\end{equation}
Thus the $n$-th line-soliton is parametrized by a pair $(k_{i_n}, k_{j_n})$
of distinct phase parameters with $\, 1 \leq i_n < j_n \leq 2N$, or 
equivalently, by the index pair~$[i_n,j_n]$. 

The Y-junction solution found by Miles~\cite{JFM1977v79p171} is, perhaps the earliest 
evidence of resonant structure present in the line-soliton solutions of KPII.
Subsequently, this solution was reconstructed using different algebraic methods
in several earlier works (see e.g., ~\cite{NewellRedekopp,Freeman,JPSJ1983v52p749}),
and more general types of resonant line-soliton solutions of KPII were reported
in some recent works including Refs.~\cite{medina,jphysa36p10519,pashaev}.
The properties of the general line-soliton solutions 
were systematically investigated in Ref.~\cite{BC} where these solutions were
characterized by developing an asymptotic analysis of the $\tau$-function.
The special case of the $N$-soliton solutions was extensively studied
in Ref.~\cite{Kodama}. In particular, an explicit characterization of the
$N$-soliton solutions space in terms of Grassmannian $Gr(N,2N)$ was
presented for the first time in \cite{Kodama}. In this article, we extend the
work in Ref.~\cite{BC} by giving a combinatorial description of the general
line-soliton solutions of KPII, and show that the class of the general line-soliton 
solutions can be enumerated employing these combinatorial properties.
We also make remarks on those solutions in terms of the positive Grassmann cells.
Furthermore, we show that the solution space is divided into dual sub-classes
of $(N,M-N)$- and $(M-N,N)$-soliton solutions 
under the action of space-time inversion $(x,y,t) \to (-x,-y,-t)$, and give a
combinatorial interpretation of this discrete symmetry of the KPII equation.
Next we consider the $N$-soliton solutions
and describe how to construct such solutions starting only from the physical
data set of $N$ amplitudes and $N$ directions of the associated line-solitons.
We show that there exists a one-to-one correspondence between $N$-soliton solution 
space and the set of all fixed-point free involutions of the permutation group
of $2N$ elements. We exhibit how the combinatorial properties of the $N$-soliton
solutions provide a further refinement of the $N$-soliton solution space, and thus 
recover the results of Ref.~\cite{Kodama}.

\section{The KPII $\tau$-function and asymptotic line-solitons}
\label{s:general}

In this section, we investigate the general properties and
asymptotic behavior of the $\tau$-function associated with the
general line-soliton solutions of the KPII equation.
As before, we consider the Wronskian form of the $\tau$-function given by
Eq. \eqref{e:tau} where the functions $\{f_n\}_{n=1}^N$ are linear
combinations of exponentials with $M$ distinct phases as in
Eq. \eqref{e:f}. Furthermore, we can assume without loss of generality,
that the phase parameters are ordered as $k_1<k_2<\ldots<k_M$.

\subsection{Properties of the $\tau$-function}
The Wronskian in Eq.~\eqref{e:tau} can be expressed as
\begin{equation}
\tau(x,y,t)= \det(A\,\Theta\,K)\,,
\label{e:taudet}
\end{equation}
where $A=(a_{nm})$ is the $N\times M$ coefficient matrix,
$\Theta= \diag(e^{\theta_1}, \ldots, e^{\theta_M})$, and
 the $M\times N$ matrix $K$ is given by $K=(k_m^{n-1}), \,
m=1,2, \ldots, M, \, n=1,2,\ldots,N$. Expanding the determinant
in Eq.~\eqref{e:taudet} using Binet-Cauchy formula yields the
following explicit form of the $\tau$-function 
\begin{equation}
\tau(x,y,t)= \sum_{1\le m_1<\dots<m_N\le M}
  A(m_1,\dots,m_N)
  \,\,\,
  \exp[\,\, \theta(m_1,\ldots,m_N) \,]
  \!\!\prod_{1\le s < r\le N}(k_{m_{r}}-k_{m_{s}})\,,
\label{e:tauexp}
\end{equation}
where $\theta(m_1,\ldots,m_N):=\theta_{m_1}+\theta_{m_2}+\ldots+\theta_{m_N}$,
$A(m_1,\dots,m_N)$ is the $N\times N$ minor of $A$
obtained from columns $1\le m_1<\dots<m_N\le M$, and the product
term is the Van~der~Monde determinant obtained using the rows 
$1\le m_1<\dots<m_N\le M$ of the matrix $K$ in Eq.~\eqref{e:taudet}. 
The basic properties of the 
$\tau$-function following from Eq. \eqref{e:tauexp} are listed below.
\begin{property}
\begin{enumerate}
\item
The $\tau$-function is a linear combination of real exponentials,
where each exponential term contains combinations of $N$ out of $M$ distinct 
phases given by $\theta(m_1,\dots,m_N)$. Any given phase combination
$\theta(m_1,\dots,m_N)$ actually appears in the $\tau$-function
if and only if the corresponding minor $A(m_1,\dots,m_N)$ is nonzero.
Thus there are {\em at most} $\binom{M}{N}$ terms in the $\tau$-function.
\item
If $M=N$, the corresponding $\tau$-function in Eq.~\eqref{e:tauexp}
contains only one exponential term which
generates the trivial solution $u(x,y,t)=0$ of KPII via Eq. \eqref{e:u}.
Hence, $M>N$ for nontrivial solutions.
\item
If rank$(A) < N$, then all the $N \times N$ minors of $A$ vanish identically, 
leading to the trivial case $\tau = 0$. Moreover, for rank$(A)=N$,
if all minors $A(m_1,\dots,m_N) \geq 0$ then
$\tau(x,y,t) > 0, \, \forall (x,y,t) \in \Real^3$. Therefore,
the resulting solution $u(x,y,t)$ of the KPII equation is non-singular.
\item
The transformation $A \rightarrow CA$ where $C \in GL(N, \Real)$
corresponds to an overall rescaling $\tau \to \det(C)\tau$,
of the $\tau$-function in Eq. \eqref{e:taudet},
which leaves the solution $u(x,y,t)$ invariant.
This $GL(N, \Real)$ freedom can be exploited to choose the coefficient
matrix $A$ in Eq. \eqref{e:f} to be in the reduced row-echelon form (RREF).
\item 
The transformation $A \rightarrow AD,\,\Theta \rightarrow D^{-1}\Theta$ 
where $D \in GL(M, \Real)$ leaves the $\tau$-function in Eq. \eqref{e:taudet}
invariant. In particular, a diagonal matrix $D$ with diagonal elements 
$d_m>0,\, m=1,\ldots,M$, leaves the functions $\{f_n\}_1^N$ in Eq.~\eqref{e:f} 
invariant by simultaneously rescaling the $m^{\mathrm th}$ column 
of $A$ by $d_m$, and shifting the each phase constant in Eq.~\eqref{e:theta}
as $\theta_{m,0} \rightarrow \theta_{m,0}-\log(d_m)$.
\item
If for any given $n,\,m$ in Eq. \eqref{e:f}, we take $f_n = e^{\theta_m}$ 
such that the $n^{\mathrm th}$
row of $A$ has only one non-zero entry $a_{nm}=1$, then
the minors $A(m_1,\dots,m_N) = 0,\, m \notin \{m_1,\dots,m_N\}$.
The resulting $\tau$-function can be expressed as
$\tau(x,y,t) = e^{\theta_m}\tau_0(x,y,t)$,
where $\tau_0(x,y,t)$ contains at most $M-1$ phases (all but
$\theta_m$) which appear in phase combinations of $N-1$ distinct phases.
It is then evident from Eq.~\eqref{e:u} that $\tau(x,y,t)$ and $\tau_0(x,y,t)$
generate the same solution of KPII. Such $\tau$-functions are {\em reducible} 
in the sense that they can be effectively obtained from a Wronskian of
$N-1$ functions with $M-1$ distinct phases.
\end{enumerate}
\label{p:tau}
\end{property}
\noindent
For the remainder of this paper, we will consider the coefficient matrix~$A$ 
to be in RREF. Furthermore, to avoid trivial and reducible cases, and to ensure 
that the solution $u(x,y,t)$ of KPII resulting from the $\tau$-function
in Eq. \eqref{e:tauexp} are non-singular, we will impose the following restrictions
on the coefficient matrix~$A$:
\begin{condition} 
\begin{enumerate}
\item
Positivity:\,\, Rank($A) = N < M$ and all nonzero minors of $A$ are positive.
\item
Irreducibility:\,\, Each column of $A$ contains at least one nonzero element, and 
each row of $A$ contains at least one nonzero element in addition to the 
pivot (first non-zero) entry.
\end{enumerate}
\label{positive}
\end{condition}
%
\begin{remark}
The matrices satisfying Condition \ref{positive}(i) above, are called 
totally non-negative (TNN) matrices. The classification of the $(N_-,N_+)$-soliton 
solutions is then given by the classification of the $N\times M$ irreducible
TNN matrices $A$ in RREF. From a more geometric perspective, each TNN matrix 
parametrizes a unique cell in the TNN Grassmannian 
$Gr^+(N,M)$ (see e.g. \cite{postnikov:06}), and the classification of the soliton solutions 
corresponds to a further refinement of the Schubert decomposition of $Gr(N,M)$ into
TNN Grassmann cells (see \cite{Kodama} for the case $M=2N$). 
The refinement is given by a classification of the coefficient matrix $A$, and
the minors $A(m_1,\ldots,m_N)$ represent the Pl\"ucker coordinates of $Gr(N,M)$.
We will discuss the geometric 
structure of this classification in a future communication~\cite{CK}.
\end{remark}

\subsection{Dominant phase combinations and asymptotic line-solitons}
\begin{figure}[t!]
\centering
\raisebox{0.87in}{(a)}\raisebox{-0.15cm}{\includegraphics[scale=0.6]{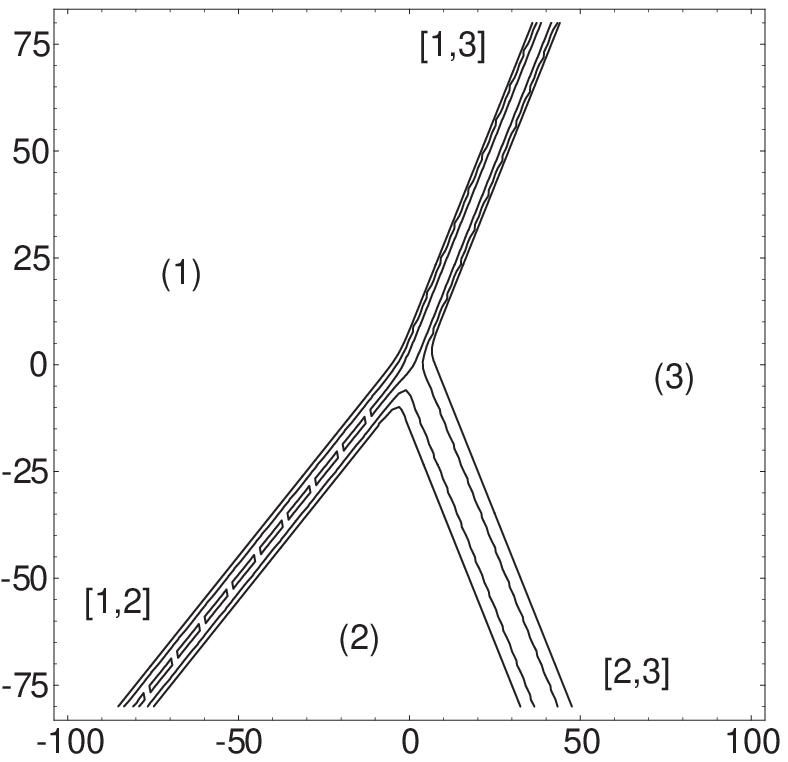}} \hskip 1.5cm
\raisebox{0.87in}{(b)}\includegraphics[scale=0.6]{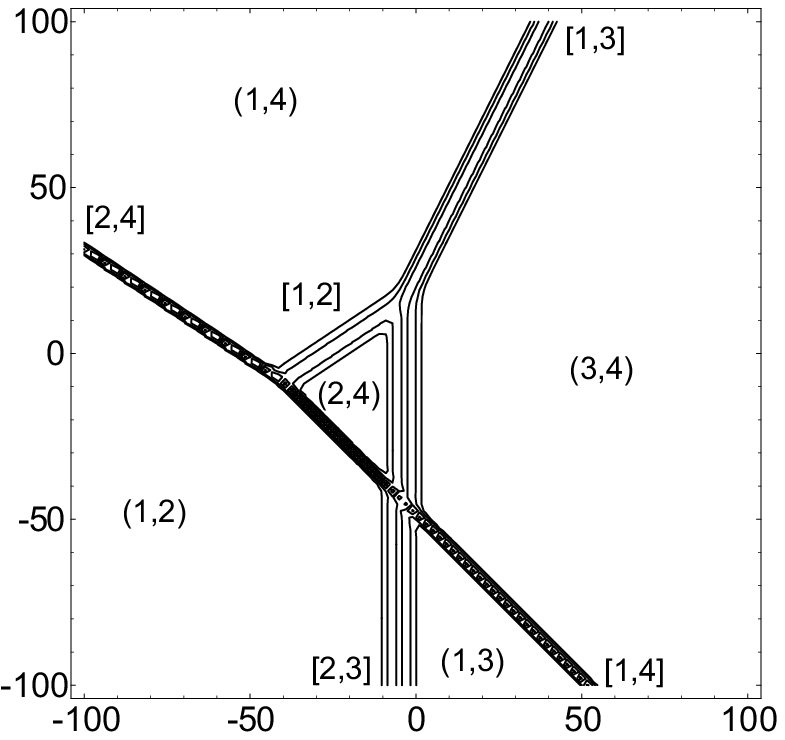} 
\caption{Dominant phase combinations in different regions
 of the $(x,y)$-plane (labeled by the indices in parentheses, i.e.
 $(i,j)=\theta(i,j)$)
 and the asymptotic line-solitons (labeled by the indices in square braces)
 for two different line-soliton solutions:
(a)~a $(2,1)$-soliton with 
   $(k_1,k_2,k_3)=(-1,0,\frac12)$
   at $t=0$; 
(b)~a $(2,2)$-soliton with $(k_1,\dots,k_4)=(-1,-\frac12,\frac12,2)$
 at $t=0$.}
\label{f:dominantphases}
\end{figure}

The spatial structure of the solution $u(x,y,t)$ is determined
from the asymptotic behavior of the $\tau$-function
in the $(x,y)$-plane and for finite values of~$t$. Notice that the $\tau$-function 
in Eq. \eqref{e:tauexp} associated with an irreducible coefficient matrix $A$ is a 
sum of real exponentials with positive coefficients. If only one phase combination 
$\theta(m_1,\dots,m_N)(x,y,t)$ in the $\tau$-function is dominant in a certain
region of the $(x,y)$-plane at a given time, then $\tau \sim exp(\theta(m_1,\dots,m_N))$.
Consequently, the solution $u(x,y,t)$ of KPII generated by the
$\tau$-function~\eqref{e:tauexp} is exponentially small at all points
in the interior of any dominant region, and is localized at the boundaries 
where a balance exists between at least two dominant 
phase combinations in the $\tau$-function~\eqref{e:tauexp}.
Such boundary is identified by the equation
$\theta(m_1,\dots,m_N) = \theta(m'_1,\dots,m'_N)$,
which defines a line segment in the $(x,y)$-plane for each $t$.
Note that this phenomenon also arises for the one-soliton
solution~\eqref{e:onesoliton},
which is localized along the line $\theta_1=\theta_2$ corresponding
to the boundary of the two regions of the $(x,y)$-plane where $\theta_1$
and $\theta_2$ dominate.
In the one-soliton case, these two regions are simply half-planes, whereas
in the general case the dominant regions could be bounded or unbounded
(see e.g. Figure~ \ref{f:dominantphases}).
A detailed analysis of the asymptotic behavior of the $\tau$-function was 
carried out in Ref.~\cite{BC}, the main results are summarized below.
\begin{proposition}
The asymptotic properties of $\tau$-function~\eqref{e:tauexp} for finite 
values of $t$, and for generic values of phase parameters
$k_1,\ldots,k_M$, are as follows:\\
(i)\,\, The dominant phase combinations of the $\tau$-function in adjacent 
regions of the $(x,y)$-plane as $y\to\pm\infty$, contain
$N-1$~common phases and differ by only a single phase.
The transition $\theta(i,m_2,\dots,m_N) \mapsto \theta(j,m_2,\dots,m_N)$
between any two such dominant phase combinations
occurs along the line defined by $[i,j]:\,\theta_i=\theta_j,\,\, i\neq j$,
where a single phase $\theta_i$ in one dominant phase combination
is replaced by a phase $\theta_j$. \\
(ii)\,\, Along the single phase transition line $[i,j]$, the dominant phase balance yields  
\begin{subequations}
\begin{equation}
\tau(x,y,t)\sim
  C_i\,e^{\theta(i,\=m_2,\dots,\=m_N)}
  + C_j\,e^{\theta(j,\=m_2,\dots,\=m_N)}
\label{e:tauasymp}
\end{equation}
asymptotically as $y\to\infty$ or as $y\to-\infty$, where
the coefficients $C_i,\, C_j$ depend on appropriate
Van~der~Monde determinants and non-vanishing minors of the coefficient 
matrix~$A$. The asymptotic behavior of the solution along $[i,j]$ is given by
\begin{equation}
u(x,y,t) \sim \half(k_j-k_i)^2\sech^2\half(\theta_j-\theta_i+\delta_{ij}) \,,
\label{e:uasymp}
\end{equation}
which defines an asymptotic line-soliton.
\end{subequations}
\\
(iii)\,\, The number of the asymptotic line-solitons is invariant in time, and
so are their amplitudes and directions.
In particular, the soliton direction is given by the normal direction 
of $[i,j]$, which is~$c_{i,j}= k_i+k_j$,
and the soliton amplitude is given by $a_{ij}=|k_i-k_j|$.
\label{P:solitons}
\end{proposition} 
In view of Proposition \ref{P:solitons}, it is natural to denote
an asymptotic line-soliton by the index pair $[i,j]$ which labels 
the asymptotic direction as well as the phase parameters in Eq. \eqref{e:uasymp}.
The asymptotic properties of the $\tau$-function reveal that there
exists a certain pairing $(k_i, k_j),\,\, 1\leq i < j \leq M$ between the phase 
parameters, which in turn defines an asymptotic line-soliton.
However, Proposition \ref{P:solitons} does not specify how to determine
these pairings in a given solutions, or equivalently, which phase combinations
are actually dominant in a given $\tau$-function as $|y| \to \infty$. It is still 
necessary to identify the particular set of asymptotic line-solitons associated 
with any given $\tau$-function of Eq.~\eqref{e:tauexp}. For this purpose, we
first need a result derived in Ref~\cite{BC} ((Lemma 3.1) regarding the dominant 
phases.
\begin{lemma}
(Dominant phase conditions)~
Along the line $[i,j]:\theta_i=\theta_j$ with $i<j$,
the phases $\theta_1,\dots,\theta_M$ satisfy the following relations,
where $\theta:=\theta_i=\theta_j$, below.
\begin{enumerate}
\par\kern-0.5\medskipamount
\itemsep 0pt
\parsep 0pt
\item 
As $y\to\infty$,\,
$\theta_m<\theta ,\, \forall \, m\in\{i+1,\dots,j-1\}$,
and $\theta_m>\theta ,\, \forall \, m\in\{1,\dots,i-1,j+1,\dots,M\}$;
\item
as $y\to-\infty$,\,
$\theta_m>\theta,\, \forall \, m\in\{i+1,\dots,j-1\}$,
and $\theta_m<\theta,\, \forall \, m\in\{1,\dots,i-1,j+1,\dots,M\}$.
\end{enumerate}
\label{dominantphase}
\end{lemma}
The proof of this Lemma based on the ordering $k_1<\cdots<k_M$ is
an easy exercise for the reader.
Lemma \ref{dominantphase} provides a simple yet useful way to determine
the dominant phase combinations along the line $[i,j]$. However, 
a given phase combination $\theta(m_1,\ldots,m_N)$ can only be dominant if 
it is in fact, {\em present} in the $\tau$-function of Eq.~\eqref{e:tauexp}, i.e.,
if the corresponding coefficient minor $A(m_1,\ldots,m_N) \neq 0$ in the 
$\tau$-function. Therefore, in order to obtain a complete characterization of the
asymptotic line-solitons it is necessary to consider the structure
of the $N \times M$ coefficient matrix $A$ in addition to
Lemma \ref{dominantphase}.
Each asymptotic line-soliton $[i,j]$ of a
$(N_-,N_+)$-soliton solution of KPII are uniquely determined by a pair of
columns of the coefficient matrix $A$, 
as prescribed below. Once again, the details can be found in Ref.~\cite{BC}.
\begin{proposition}
\label{P:pairing}
The $(N_-,N_+)$-soliton
solution of KPII generated from the $\tau$-function
in Eq.~\eqref{e:tauexp} has exactly $N_+= N$ asymptotic line-solitons
as $y\to\infty$ and $N_-=M-N$ asymptotic line-solitons as $y\to-\infty$.
The necessary and sufficient conditions for an index pair $[i,j]$
to identify an asymptotic line-soliton are determined by the ranks of two 
sub-matrices of $A$ defined below in terms of their column indices 
\begin{equation*}
X[ij] := \left[ 1,2,\ldots, i-1, j+1, \ldots, M \right] \qquad
Y[ij] := \left[i+1, \ldots j-1 \right] \,.
\end{equation*}
The rank conditions are then as follows:\\
(i)\, Each asymptotic line-soliton as $y\to\infty$ is labeled by 
a unique index pair $[e_n,j_n]$ with $e_n < j_n$ where
$\{e_n\}_{n=1}^N$ label the pivot columns of $A$.
Moreover, if $\rank(X[e_nj_n]):=r_n$, then 
$$r_n \le N-1 \qquad and \qquad \rank(X[e_nj_n]|e_n) \,\, = \,\,\rank(X[e_nj_n]|j_n) \,\,
= \,\, \rank(X[e_nj_n]|e_n,j_n) \,\,= \,\,r_n+1 \,.$$
(ii)\, An asymptotic line-soliton as $y\to -\infty$ is labeled
by a unique index pair $[i_n,g_n]$ with $i_n < g_n$ where
$\{g_n\}_{n=1}^{M-N}$ label the non-pivot columns of $A$.
Moreover, if $\rank(Y[i_ng_n]) := s_n$, then 
$$ s_n \le N-1 \qquad and \qquad \rank(Y[i_ng_n]|i_n) \,\, = \,\, \rank(Y[i_ng_n]|g_n)
\,\, = \,\,\rank(Y[i_ng_n]|i_n,g_n)\,\,= \,\,s_n+1 \,.$$
Above, $(Z|m,n)$ denotes the sub-matrix~$Z$ of $A$ augmented by the
columns $m$ and $n$ of $A$.
\end{proposition}
Given the $\tau$-function data, which consist of $M$ distinct phase parameters
$k_1,\ldots,k_M$ and a matrix $A$ satisfying Condition \ref{positive},
Propositions \ref{P:solitons} and \ref{P:pairing} provide an explicit
way to identify all the asymptotic line-solitons of the corresponding
solution of the KPII equation. This method is illustrated via the
examples below.
\begin{example}
Figure~\ref{f:dominantphases}(a) illustrates a $(2,1)$-soliton 
Y-junction solution \cite{JFM1977v79p171} describing the resonant interaction
of two line-solitons mentioned in Section \ref{s:introduction} 
(see Figure~\ref{f:kpfig}(c)).
This solution corresponds to $N=1,\, M=3$, and is generated by the
$\tau$-function and the coefficient matrix $A$,
$$
\tau(x,y,t)= e^{\theta_1}+e^{\theta_2}+e^{\theta_3}\,, \qquad \qquad 
A= \begin{pmatrix}
 1 &1 &1 
\end{pmatrix} \,.
$$
In this case we know from Proposition \ref{P:pairing} that the number
of asymptotic line-solitons as $y\to \infty$ and as $y\to -\infty$ are 
one and two, respectively. Applying the rank conditions from 
Proposition \ref{P:pairing}(i) to the 
pivot column $e_1=1$, we see that for the soliton $[1,j_1]$ as 
$y \to \infty$, $\rank(X[1j_1]) = 0$ since $N=1$. Hence, $j_1=3$, 
so that the line-soliton is $[1,3]$, which (according to Proposition
\ref{P:solitons}) corresponds to the
dominant balance of the phases $\theta_1$ and $\theta_3$ in the $\tau$-function.
As $y \to -\infty$, Proposition \ref{P:pairing}(ii) implies that for
the non-pivot columns $g_1=2$ and $g_2=3$, we should have 
$\rank(Y[i_12])=\rank(Y[i_23]) = 0$ as well. Consequently, $i_1=1, \, i_2=2$,
and the resulting line-solitons $[1,2],\, [2,3]$ correspond to the dominant
balance of the phase pairs $(\theta_1,\theta_2)$ and $(\theta_2,\,\theta_3)$,
respectively.  Thus, the $(x,y)$-plane is partitioned in
three disjoint regions where each of the phases $\theta_1,\, \theta_2$
and $\theta_3$ dominates, and the solution is localized along
the phase transition lines which mark the asymptotic line-solitons.
\label{E:2to1}
\end{example}
\begin{example}
Consider $N=2$ and $M=4$ corresponding to a $(2,2)$-soliton solution 
as shown in Figure~\ref{f:dominantphases}(b), and generated by the 
$\tau$-function in Eq.~\eqref{e:tau} with 
$$
f_1 = e^{\theta_1}-e^{\theta_4}, \quad
f_2 = e^{\theta_2}+e^{\theta_3}+ e^{\theta_4}, \qquad \qquad
A= \begin{pmatrix}
1 &0 &0 &\!-1 \\ 0 &1 &1 &1 
\end{pmatrix} \,.
$$
The pivot columns of~$A$ are labeled by the indices $\{e_1,e_2\}=\{1,2\}$,
and the non-pivot columns by the indices $\{g_1,g_2\}=\{3,4\}$.
According to Proposition~\ref{P:pairing}, the number of asymptotic line
solitons are $N_+=N_-=2$. They are identified by the index pairs $[1,j_1],\, [2,j_2]$ 
as $y\to \infty$, for some $j_1>1$ and $j_2>2$; and by the index pairs $[i_1,3]\,,[i_2,4]$ 
as $y\to -\infty$, for some $i_1<3$ and $i_2<4$.
We first determine the asymptotic line-solitons as $y\to\infty$
using the rank conditions prescribed in Proposition~\ref{P:pairing}(i).
For the first pivot column $e_1=1$, starting from $j=2$ and then incrementing 
the value of $j$ by one, we check the rank of each sub-matrix $X[1j]$.
Proceeding in this way, we find that the rank conditions are satisfied 
when $j=3$: $X[13]= \bigl(\begin{smallmatrix}\!-1\\1\end{smallmatrix}\bigr)$.
So $\rank(X[13])=1=N-1$. Moreover, $\rank(X[13]|1)= \rank(X[13]|3)=
\rank(X[13]|1,3) =2$. Thus, the first asymptotic line-soliton
as $y\to\infty$ is identified by the index pair $[1,3]$.
For $e_2=2$, proceeding in a similar manner as above we
find that $j=3$ does not satisfy the rank conditions (since
$X[23]$ has rank~2) but $j=4$ does.
Therefore, the asymptotic line-solitons as $y\to\infty$ are given by
the index pairs $[1,3]$ and $[2,4]$.

Then we consider the asymptotics for $y\to-\infty$.
Starting with the non-pivot column $g_1=3$, we apply the rank conditions
in Proposition~\ref{P:pairing}(ii) to the column $i=2$.
Then, we have $Y[23]=\emptyset$, and
$\rank(Y[23]|2)= \rank(Y[23]|3)= \rank(Y[23]|2,3)=1$.
Hence, the pair $[2,3]$ identifies an asymptotic line-soliton
as $y\to-\infty$. For $g_2=4$, we consider $i=1,2,3$ and find that
the rank conditions are satisfied only for $i=1$. In this case,
$Y[14]= \bigl(\begin{smallmatrix}0&0\\1&1\end{smallmatrix}\bigr)$,
so $\rank(Y[14])= 1=N-1$ and
$\rank(Y[14]|1)= \rank(Y[14]|4)= \rank(Y[24]|1,4)=2$.
Thus, the index pair $[1,4]$ identifies the other
asymptotic line-soliton as $y\to-\infty$.

The dominant phase regions in the $(x,y)$-plane for this $(2,2)$-soliton
solution can be also identified from Proposition \ref{P:solitons}.
First note that the dominant phase combination along $x \to -\infty$ 
is given by $\theta(1,2)=\theta_1+\theta_2$ for finite $y$. This follows from
Eq.~\eqref{e:theta} and the ordering $k_1<k_2<k_3<k_4$ of the phase parameters.
As the slope of the normal to the transition line $\theta_i=\theta_j$,
given by the direction $c_{ij}=k_i+k_j$ decreases 
from negative $x$-axis to positive $x$-axis for $y>0$, the asymptotic line-solitons
are sorted clockwise as $[2,4]$ and $[1,3]$ as $y \to \infty$.
Thus the dominant phase combinations associated with single phase transitions 
(cf. Proposition \ref{P:solitons}) as $y \to \infty$ are given by
\begin{equation*}
\theta(1,2) \mapto24 \theta(1,4) \mapto13 \theta(3,4) \,.
\end{equation*}
When $y<0$, the soliton direction parameter $c_{ij}$ increases from 
negative $x$-axis to positive $x$-axis. Consequently, the asymptotic line 
solitons as $y \to -\infty$ are sorted counter-clockwise as $[2,3]$ and $[1,4]$,
and determine the dominant phase combinations in the $(x,y)$-plane for $y\to -\infty$ as
follows:
\begin{equation*}
\theta(1,2) \mapto23 \theta(1,3) \mapto14 \theta(3,4) \,.
\end{equation*}
The asymptotic line-solitons and dominant phase combinations are
shown in Figure~\ref{f:dominantphases} where the phase parameters
are chosen such that $c_{14}=1>c_{23}=0$. Note that in addition to 
the unbounded dominant regions, there is also a bounded region in the
$xy$-plane where $\theta(2,4)$ is the dominant phase combination.
The boundaries of this region is formed by the line-solitons [1,4] and 
[2,3] as $y \to -\infty$, together with the intermediate line-soliton [1,2].
\label{E:2to2}
\end{example}
The above examples illustrate how to identify the asymptotic line
solitons and the dominant phase combinations of a $(N_-,N_+)$-soliton 
solution of KPII from the $\tau$-function in an algorithmic fashion.
First, apply the rank conditions given in Proposition \ref{P:pairing} to each 
pivot column and to each non-pivot column of the coefficient matrix $A$ to 
identify the asymptotic line-solitons as $y \to \pm \infty$.
Next, note that for a $\tau$-function in Eq.\eqref{e:tauexp} associated with a 
coefficient matrix $A$ in RREF, the dominant phase combination as
$x \to -\infty$ is uniquely given by $\theta(e_1,\ldots,e_N)$.
This is due to the fact that the phase parameters are ordered as 
$k_1<k_2<\ldots<k_M$, and the coefficient $A(e_1,\ldots,e_N)$ of the  
term $e^{\theta(e_1,\ldots,e_N)}$ being the minor of pivot columns, is 
lexicographically the first non-vanishing minor of $A$ with $A(e_1,\ldots,e_N)=1$.
Since the line-solitons are sorted according to their direction parameter $c_{ij}$,
the dominant phase combinations in the $(x,y)$-plane can be then determined
from Proposition \ref{P:solitons} starting from the dominant phase combination 
$\theta(e_1,\ldots,e_N)$ as $x \to -\infty$. 
\subsection{Index pairing and derangements}
\begin{figure}[h!]
\centering
\includegraphics[scale=0.45]{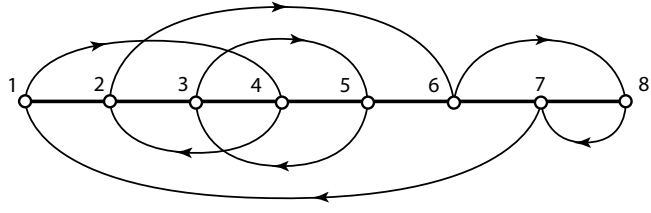} 
\caption{The open chord diagram for $\pi=(46523817)$.}
\label{f:chord10}
\end{figure}

We show here that 
the set of unique index pairings in Proposition \ref{P:pairing}, 
$\{[e_n,j_n]\}_{n=1}^N \cup \{[i_n,g_n]\}_{n=1}^{M-N}$ identifying the asymptotic 
line-solitons as $|y| \to \infty$, has a combinatorial interpretation.
Let $[M] := \{1,2,\ldots,M\}$ be the integer set and recall that
$\{e_1,\ldots,e_N\} \cup \{g_1,\ldots,g_{M-N}\}$ is a disjoint partition
of $[M]$. Define the pairing map $\pi: [M] \to [M]$ 
according to Proposition \ref{P:pairing}(i) \& (ii) as
\begin{equation}
\pi(e_n) = j_n\,, \,\, n=1,2,\ldots,N\,, \qquad
\pi(g_n) = i_n\,, \,\, n=1,2,\ldots,M-N\,,
\label{e:pair}
\end{equation}
where $e_n$ and $g_n$ are respectively, the pivot and non-pivot indices of
the coefficient matrix $A$. Then one can show that 
the map $\pi: [M] \to [M]$ is a bijection, that is, $\pi$ is a {\em permutation} of the set $[M]$. It is 
sufficient to show that the image $\pi([M])$ is a set of distinct elements.
First assume the contrary, i.e., suppose $\pi(l)=\pi(l')=m$ for two distinct
elements $l,l' \in [M]$. Then consider the dominant phase combinations 
associated with the single phase transitions (see 
Proposition \ref{P:solitons}(i)) starting with the 
$\theta(e_1,\ldots,e_N)$ as $x \to -\infty$, proceeding
clockwise, and finally back to $\theta(e_1,\ldots,e_N)$ after a complete
revolution. There are altogether $M$ such single phase 
transitions: $N$ transitions as $y \to \infty$ corresponding to the pivot 
indices $\{e_1,\ldots,e_N\}$, and $M-N$ transitions as $y \to -\infty$ corresponding 
to the non-pivot indices $\{g_1,\ldots,g_{M-N}\}$.
Without loss of generality, if we assume that the
transition $l \to m$ occurs before $l' \to m$, then there
must be an intermediate $m \to m'$ transition in between those two single phase 
transitions. Now if $\theta_m \in \theta(e_1,\ldots,e_N)$ then the transition
$m \to m'$ must occur before the $l \to m$ transition can take place.
Consequently, the intermediate $m \to m'$ transition can {\em not} occur as
each transition occurs only once during a complete revolution in the
$(x,y)$-plane. On the other hand, if $\theta_m \notin \theta(e_1,\ldots,e_N)$ 
then the $m \to m'$ transition must
occur after the $l' \to m$ transition, and again the intermediate $m \to m'$ 
transition can not take place. Either way, we reach a contradiction implying
that $\pi: [M] \to [M]$ is one-to-one, and therefore a bijection, thus proving our claim.
Note also that $\pi$ has no fixed point because $\pi(e_n) > e_n,\, n=1,\ldots,N$ 
and $\pi(g_n) < g_n, \, n=1,\ldots,M-N$. A permutation $\pi$ with no fixed
point is called a {\em derangement}, and an element $l \in [M]$ is called an
{\em excedance} of $\pi$ if $\pi(l) > l$. We can summarize the above discussions
as follows.
\begin{proposition}
\label{P:derangement}
The pairing map $\pi$ defined by Eq.~\eqref{e:pair} is a derangement of $[M]$
with $N$ excedances which are given by the pivot indices $\{e_1,\ldots,e_N\}$
of the coefficient matrix $A$ in RREF.
\end{proposition}
In Example \ref{E:2to1}, the pivot index $e_1=1$, and the non-pivot indices
$g_1=2,\, g_2=3$ for the coefficient matrix $A$, with $M=3$. The soliton pairings 
are:\, $[1,3]$ as $y \to \infty$,
and $[1,2],\, [2,3]$ as $y \to -\infty$. The corresponding pairing map from 
Eq.~\eqref{e:pair} is given by
\[
\pi= \begin{pmatrix}
1 &2 &3 \\ 3 &1 &2
\end{pmatrix}
\]
in the bi-word notation of permutation, and $\pi$ has only one excedance:\, $\{1\}$.
In Example \ref{E:2to2}, $M=4$, the pivot and non-pivot indices are 
$\{e_1=1,\, e_2=2\}$ and $\{g_1=3,\, g_2=4\}$, respectively. The pairing map
\[
\pi= \begin{pmatrix}
1 &2 &3 &4\\ 3 &4 &2&1
\end{pmatrix}
\]
corresponds to the line-solitons $[1,3],\,[2,4]$ as $y \to \infty$,
and $[2,3],\, [1,4]$ as $y \to -\infty$ with excedance set $\{1,2\}$.
The pairing map $\pi$ can also be represented by an open chord diagram 
associated with permutations (see \cite{corteel:07}).
For example, shown in Figure~\ref{f:chord10} is the diagram for $\pi=(46523817)$ 
(using one-line notation for permutations).

\noindent
The four upper chords in Figure~\ref{f:chord10} originate from the 
4 excedances $\{1,2,3,6\}$ with
the right arrows indicating the increasing order of the numbers. 
The upper chords correspond to the asymptotic line-solitons for
$y\to\infty$ with the pairings $[1,4], [2,6], [3,5], [6,8]$. Similarly, the 
lower chords with reversed arrows indicate the line-solitons for $y\to-\infty$
namely, $[1,7], [2,4], [3,5], [7,8]$.
Thus, this chord diagram represents a $(4,4)$-soliton solution
of KPII. Note here that the $[3,5]$-soliton appears both in
$y\to\pm\infty$ which corresponds to a 2-cycle $(35)$ in the permutation $\pi$.

Proposition \ref{P:derangement} furnishes a combinatorial characterization of
the $(N_-,N_+)$-soliton solutions of KPII as explained below.
\begin{definition}
Let $S_+ := \{[e_n,j_n]\}_{n=1}^N$ and $S_- := \{[i_n,g_n]\}_{n=1}^{M-N}$ denote
the index sets labeling the asymptotic line-solitons as $y \to \infty$ and
as $y \to -\infty$, respectively. Then two $(N_-,N_+)$-soliton solutions of KPII are 
defined to be in the same equivalence class if their asymptotic line-solitons are labeled
by the identical sets $S_\pm$ of index pairs, where $|S_+| := N_+=N$ and 
$|S_-| := N_- = M-N$. 
\label{D:equiv}
\end{definition}
Each equivalence class of $(N_-,N_+)$-soliton solutions of KPII is uniquely
determined by a derangement $\pi$ as defined in Proposition \ref{P:derangement}.
Therefore, each equivalence class of soliton solutions is also
associated with a unique open chord diagram with $N_-=N$ upper chords
and $N_+=M-N$ lower chords. Furthermore, in view of Remark 2.3,
each derangement of $[M]$ gives a unique parametrization of a TNN Grassmann cell
in $Gr^+ (N,M)$, denoted by $W(\pi)$, and whose dimension can be computed from the
number of crossings of the corresponding chord diagram~\cite{williams:05,postnikov:06}.
In particular, the dimension of the cell $W(\pi)$ associated with a TNN matrix satisfying
Condition~\ref{positive} is given by~\cite{CK}
\[
{\rm dim}\,W(\pi)=N+C_+(\pi)+C_-(\pi)\,,
\]
where the number of crossings $C_{\pm} (\pi)$ in the chord diagram are 
defined by~\cite{corteel:07}
\begin{equation}
C_{\pm}(\pi)=\sum_{i=1}^MC_{\pm}(i)\,, \quad{\rm where}\quad\left\{\begin{array}{llll}   
C_+(i):=\{j:j<i<\pi(j)<\pi(i)\}\,, \\
C_-(i):=\{j:j>i\ge \pi(j)>\pi(i)\}\,.
\end{array}\right.
\label{cross}
\end{equation} 
(Note that the above definitions of $C_{\pm}(i)$ are switched from
those given in Ref.~\cite{corteel:07}). 
For the example in Figure~\ref{f:chord10}, the chord diagram associated 
to $\pi=(46523817)$ has $C_+(\pi)=2, C_-(\pi)=2$, and the 
corresponding Grassmann cell in $Gr^+(4,8)$ has dimension $8=4+2+2$. The
top (maximal) dimensional cell in $Gr^+(4,8)$ of dimension $16=4\times 4$
corresponds to a chord diagram with the maximum number of crossings, i.e. 
$\pi_{top}=(56781234)=(15)(26)(37)(48)$ with 
$C_+(\pi_{top})=C_-(\pi_{top})=6$, and $\dim W(\pi_{top}) = 16=4+6+6$.
An equivalence class of $(N_-,N_+)$-soliton solutions of KPII with a given
pairing map $\pi$ can thus be associated with a unique TNN Grassmann cell $W(\pi)$,
and the number of free variables parametrizing the $(N_-,N_+)$-soliton solution
space is given by $\dim W(\pi)$. We next proceed to calculate the total number
of $(N_-,N_+)$-soliton equivalence classes for given values of $M$ and $N$, and 
enumerate these equivalence classes according to the dimensions of their 
solution spaces.

Let $\mathcal{S}_M$ denote the permutation group of $[M]$, and let 
$\mathcal{D}_M \subset \mathcal{S}_M$ be the set of all derangements
of $[M]$. Then the derangements can be enumerated according to the number of
excedances $e(\pi)$ of $\pi$ by the generating polynomial
\begin{equation*}
D_M(p) = \sum_{\pi \in \mathcal{D}_M} p^{e(\pi)} = \sum_{N=1}^{M-1}D_{N,M}p^N\,, \quad M \geq 1\,,
\end{equation*}
where the coefficients $D_{N,M}$ denote the number of derangements of $[M]$ with $N$ excedances.
The total number of derangements of $\mathcal{S}_M$ is then given by $|\mathcal{D}_M|= D_M(1)$. The explicit 
formula for the derangement polynomial $D_M(p)$ is 
obtained from the exponential generating function~\cite{Roselle} with $D_0(p):=1$,
\begin{equation}
D(p,z) = \sum_{M=0}^{\infty} D_M(p)\frac{z^M}{M!} = \frac{1-p}{e^{zp}-pe^z}
\label{e:dpz}
\end{equation}
The first few polynomials are given by 
$D_1(p)=0,\,\, D_2(p)=p,\,\, D_3(p)=p+p^2,\,\, D_4(p) = p+7p^2+p^3$.
Moreover, the polynomials $D_M(p)$ are {\em symmetric}, that is, its 
coefficients satisfy 
\begin{equation}
D_{N,M} = D_{M-N,M}\,, \qquad N=1,2,\ldots,M-1\,.
\label{e:symm}
\end{equation}
This and various other properties of the derangement polynomial $D_M(p)$ can
be found in Ref.~\cite{Roselle}. Note that Eq. \eqref{e:symm} is equivalent 
to the relation 
$$ p^MD_M(p^{-1}) \,\, = \,\, D_M(p) \,,$$ 
which in turn follows from the symmetry 
$$ D(p^{-1},\, zp) \,\,  =  \,\, D(p,z) $$ 
of the exponential generating function, and can be verified directly from Eq.~\eqref{e:dpz}.

The above formulas give the number $D_{N,M}$ of equivalence classes for
the $(N_-,N_+)$-solitons
of KPII for a given $M=N_-+N_=$ and $N=N_+$. Thus, when $M=3$, there are only 2 classes of line-soliton
solutions namely, the $(2,1)$-soliton as in Example \ref{E:2to1}, and also $(1,2)$-solitons
which are related to the $(2,1)$-solitons by the inversion symmetry:\,
$(x,y,t) \to (-x,-y,-t)$. When $M=4$, there are one type each of the $(3,1)$- and
$(1,3)$-soliton solutions related via the inversion symmetry, but also 7 distinct
types of $(2,2)$-soliton solutions. It is also possible to obtain a further
refinement of the total number of soliton equivalence classes by introducing
a $q$-analog of the derangement number $D_{N,M}$ namely,
$$ D_{N,M}(q) = \sum_{r=N} ^{N(M-N)}D_{r,N,M}\, q^r \,, $$
where $D_{r,N,M}$ is the number of derangements of $[M]$ with $N$ excedances, and
$r-N$ crossings as defined in Eq.~\eqref{cross}. Then, $D_{N,M}(q)$ is the
generating polynomial for the Grassmann cells in $Gr^+(N,M)$ corresponding
to the irreducible TNN matrices satisfying Condition~\ref{positive}, and
$D_{r,N,M}$ is the number of those TNN cells of dimension $r$. The upper limit $N(M-N)$ in the sum
gives the dimension of the top cell in $Gr^+(N,M)$.
Equivalently, $D_{r,N,M}$ then gives the number
of $(N_-,N_+)$-soliton equivalence classes with 
$N_+=N,\, N_-=M-N$, and with $r$ free parameters. Moreover, the total
number of $(N_-,N_+)$-soliton equivalence classes is given by 
$D_{N,M}(q=1) = D_{N,M}$. 

It is interesting to note that $D_{N,M}(q)$ is related to a $q$-analog of 
the Eulerian number ~\cite{williams:05}, 
$$ 
E_{k,n}(q)=q^{n-k^2-k}\sum_{i=0}^{k-1}(-1)^i\,[k-i]_q^n\,\,q^{ki}\,
\left(\binom{n}{i}q^{k-i}+\binom{n}{i-1}\right)\,,
$$
where $[k]_q := 1+q+q^2+\ldots+q^{k-1}$ is the $q$-analog of the number $k$.
The polynomial $E_{k,n}(q)$ was recently introduced in Ref.~\cite{williams:05} 
where a rank generating function for the cells in $Gr^+(N,M)$ was derived by building
on the work of Ref.~\cite{postnikov:06}. It follows from Refs.~\cite{williams:05,corteel:07} 
that the coefficient of $q^r$ in $E_{N,M}(q)$
is the number of permutations of $[M]$ with $N$ {\em weak} exedances, and whose
chord diagrams have $r-N$ crossings as defined in Eq.~\eqref{cross}. Note that
$l \in [M]$ is called a {\em weak} exedance of a permutation $\pi \in \mathcal{S}_M$
if $\pi(l) \geq l$. The following result gives the relation between the
polynomials $D_{N,M}(q)$ and $E_{N,M}(q)$.
\begin{proposition}
For fixed $N_+=N,\, N_-=M-N$, the generating polynomial for the $(N_-,N_+)$-soliton 
equivalence classes according to the dimension of their solution spaces is
given by
$$ D_{N,M}(q)=\sum_{j=0}^{N-1}(-1)^j\binom{M}{j}\,E_{N-j,M-j}(q)\,, $$
where $E_{k,n}(q)$ are the Eulerian polynomials defined above.
\label{P:dMNq}
\end{proposition}
\begin{proof}
Let $E(N,M)$ denote the set of all permutations of $[M]$ with $N$ weak exedances,
and let $D(N,M) \subset \mathcal{D}_M$ denote the derangements of $[M]$ with $N$ exedances.
Then the polynomials $E_{N,M}(q)$ and $D_{N,M}(q)$ are given in terms of the number 
of crossings $c(\pi)$ of the permutation $\pi$ as
$$ 
E_{N,M}(q) = q^N\hspace{-0.05 in}\sum_{\pi \in E(N,M)} \hspace{-0.1 in} q^{c(\pi)} \,, 
\quad \qquad
D_{N,M}(q) = q^N \hspace{-0.05 in}\sum_{\pi \in D(N,M)}\hspace{-0.1 in} q^{c(\pi)} \,.
$$
Note that for each $n \leq N-1$, an element of $E(N,M)$ can be obtained by adding 
$n$ fixed points to the corresponding element of the derangement set $D(N-n,M-n)$.
Then for $N'=N-n, \, M'=M-n$, 
$$ 
E_{N,M}(q) =  q^N \hspace{-0.05 in}\sum_{\genfrac{}{}{0pt}{1}{S \subset [M],}{|S|=n}} 
\sum_{\pi \in D(N',M')} \hspace{-0.1 in}q^{c(\pi)} \;\; = \;\; 
\sum_{n=0}^{N-1}\binom{M}{n}D_{N-n,M-n}(q) \,. $$
Inverting the above formula, yields the desired result.  
\end{proof}
Proposition~\ref{P:dMNq} provides an explicit formula for enumerating the 
$(N_+,N_-)$-soliton equivalence classes according to the dimensions of
the associated Grassmann cells in $Gr^+(N,M)$. For example, when $M=4$ and $N=2$, 
Proposition~\ref{P:dMNq} yields $D_{2,4}(q)= q^4+4q^3+2q^2$. This implies that there 
are one cell of dimension 4 (top cell), four cells of dimension 3, and
two cells of dimension 2. The $Gr^+(2,4)$ case will be discussed in the
next section. 

It turns out that the line-soliton solutions
of KPII possess several other combinatorial properties which play significant 
roles in their classification scheme. Some of these properties were 
addressed in Ref.~\cite{Kodama} (see also ~\cite{BCa}). In this article, we
will present these combinatorial structures underlying the line-soliton 
solutions from an algebraic perspective in Section~\ref{s:elastic}.

\section{$(2,2)$-soliton solutions}
\label{s:twosolitons}
In this section we study the line-soliton solutions of KPII which
admit a pair of asymptotic line-solitons as $|y| \to \infty$. But
in general, the pair of line-solitons as $y \to \infty$ differ from
those as $y \to -\infty$ in their amplitudes and directions. We call
these the $(2,2)$-soliton solutions, which include the 2-soliton 
solutions as well.
Based on our asymptotic results of Section~\ref{s:general},
these solutions are specified by prescribing the
distinct phase parameters $\{k_1,\dots,k_4\}$ and the $2\times 4$
coefficient matrix~$A$. The $\tau$-function of any $(2,2)$-soliton 
is given by 
\begin{equation}
\tau(x,y,t)= \sum_{1\le r<s\le4} 
  (k_s-k_r)\,A(r,s)\, e^{\theta_r+\theta_s}\,.
\label{e:tau2sol}
\end{equation}
where $A(r,s)$ denotes the $2\times2$ non-negative minors of
the matrix ~$A$.
\subsection{Classification of $(2,2)$-soliton solutions}
\begin{figure}[h]
\centering
\includegraphics[scale=0.5]{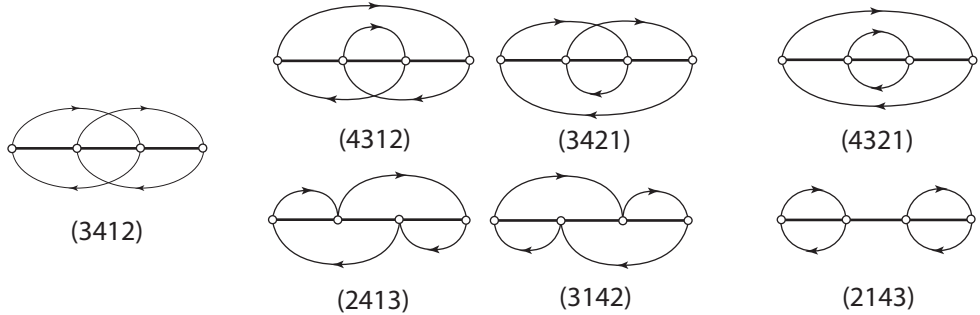} 
\caption{The open chord diagrams for the seven equivalence classes
of $(2,2)$-soliton solutions.} 
\label{f:chord2}
\end{figure}

For a given set of phase parameters $\{k_1,\dots,k_4\}$, all possible
equivalence classes of $(2,2)$-soliton solutions can be completely 
enumerated by the derangements of the index set $[4]$ with 2 excedances. 
Recall from Section~\ref{s:general}.3 that
there are altogether 7 distinct equivalence classes of $(2,2)$-soliton solutions
given by the coefficient $D_{2,4}$ of $p^2$, in the derangement polynomial $D_4(p)$.
The 7 equivalence classes can be further enumerated by the
$q$-derangement number $D_{2,4}(q)=q^4+4q^3+2q^2$ according to the number of free 
parameters spanning the solution space of each equivalence class. The coefficients
1,4,2 in $D_{2,4}(q)$ correspond to the number of $(2,2)$-soliton equivalence 
classes spanned respectively, by 4,3 and 2 free parameters. This is illustrated in Figure~\ref{f:chord2}
below by the open chord diagrams associated with the $(2,2)$-soliton solutions.

\noindent
The number sequence below each diagram in Figure~\ref{f:chord2}
is the one-line notation for the
corresponding permutation $\pi$ which also designates a certain TNN cell
$W(\pi)$ in $Gr(2,4)$. The left diagram with 2 crossings corresponds to the top
cell of dimension $4$; the middle four diagrams with 1 crossing in each correspond 
to the dimension 3 cells; while the last two diagrams have no crossing,
and correspond to the cells of dimension 2 in $Gr^+(2,4)$. We verify these
combinatorial results below by directly analyzing the $(2,2)$-soliton coefficient
matrix $A$ which also represents the TNN cells in $Gr^+(2,4)$.

The $(2,2)$-soliton solutions arise from one of the following two 
irreducible coefficient matrices in RREF:
\begin{equation}
A_1= \begin{pmatrix}
 1 &0 &\!-a &\!-b \\
 0 &1 &c &d \\
\end{pmatrix}, \quad \text{or} \quad
A_2= \begin{pmatrix}
 1 &a &0 &\!-b \\
 0 &0 &1 &d \\
\end{pmatrix}
\label{e:A2sol}
\end{equation}
with arbitrary {\em non-negative} parameters $a,b,c,d$. We classify them
according to the number of independent positive parameters in the coefficient
matrix $A_1$ or in $A_2$. Note that the matrices in Eqs.~\eqref{e:A2sol}
identify certain Schubert cells of $Gr(2,4)$. A further refinement of those
Schubert cells is given by the TNN Grassmann cells classified below.

\noindent {\bf 4 positive parameters}:\, 
There is one such case, which corresponds to the matrix $A_1$ in Eq.~\eqref{e:A2sol}
with $A_1(34) = bc-ad >0$. Note that all six exponential terms are present 
in the $\tau$-function in Eq.~\eqref{e:tau2sol}. From Proposition \ref{P:pairing}, 
one finds that the pair of asymptotic line-solitons are the {\em same} as
$|y| \to \infty$ with index pairs $([1,3], [2,4])$.  The associated pairing map given by
the permutation is $\pi=(3412)=(13)(24)$. Thus, this case
corresponds to an equivalence class of 2-soliton solutions where each asymptotic
line-soliton as $y \to \infty$ has identical amplitude and direction to another
asymptotic line-soliton as $y \to -\infty$. The 2-soliton solutions will be 
discussed further in the next subsection.
\begin{figure}[t!]
\centering
\raisebox{1.0in}{(a)}\includegraphics[scale=0.55]{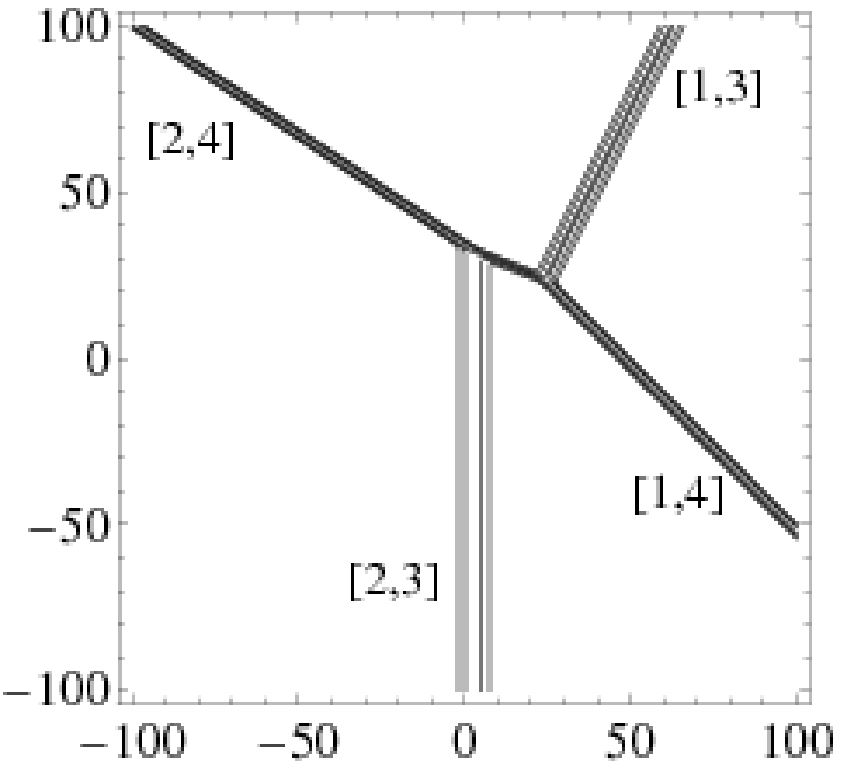}  \hskip 2cm
\raisebox{1.0in}{(b)}\includegraphics[scale=0.55]{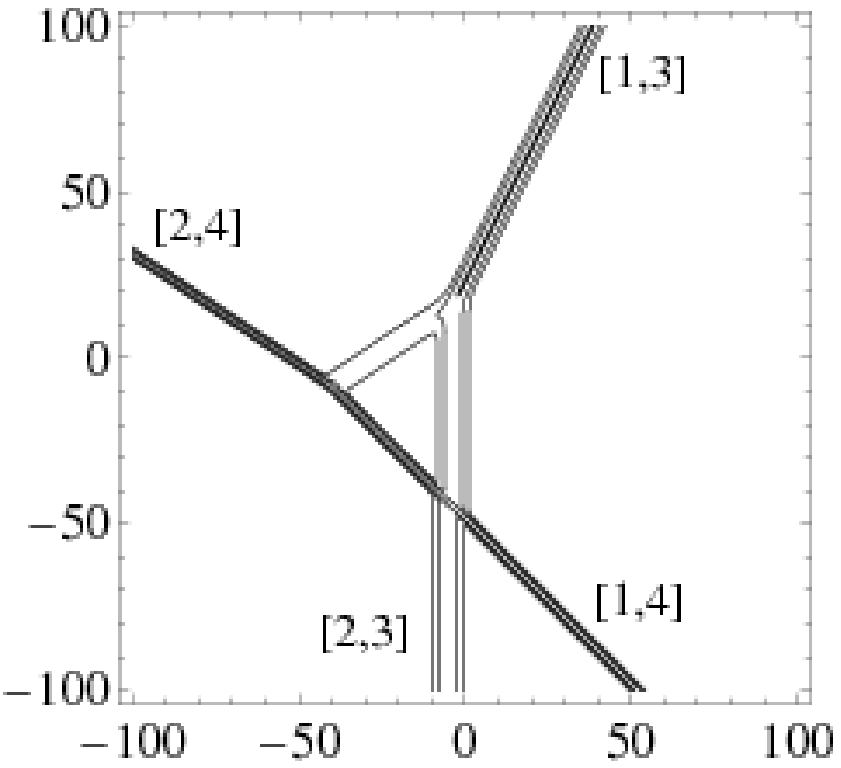} \\
\raisebox{0.5cm} {} \\
\raisebox{1.0in}{(c)}\includegraphics[scale=0.55]{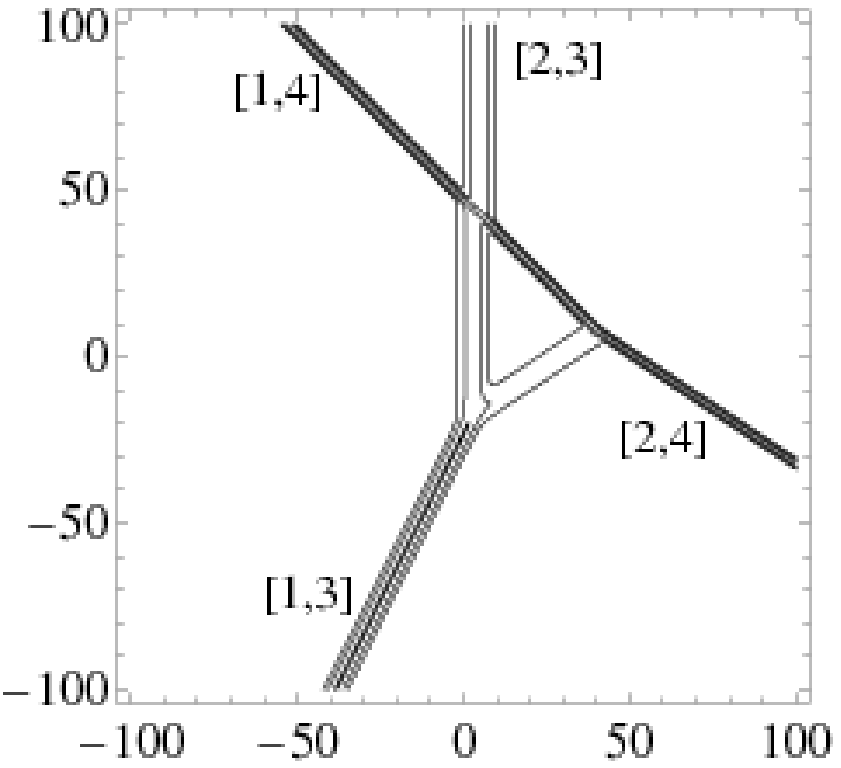}  \hskip 2cm
\raisebox{1.0in}{(d)}\includegraphics[scale=0.55]{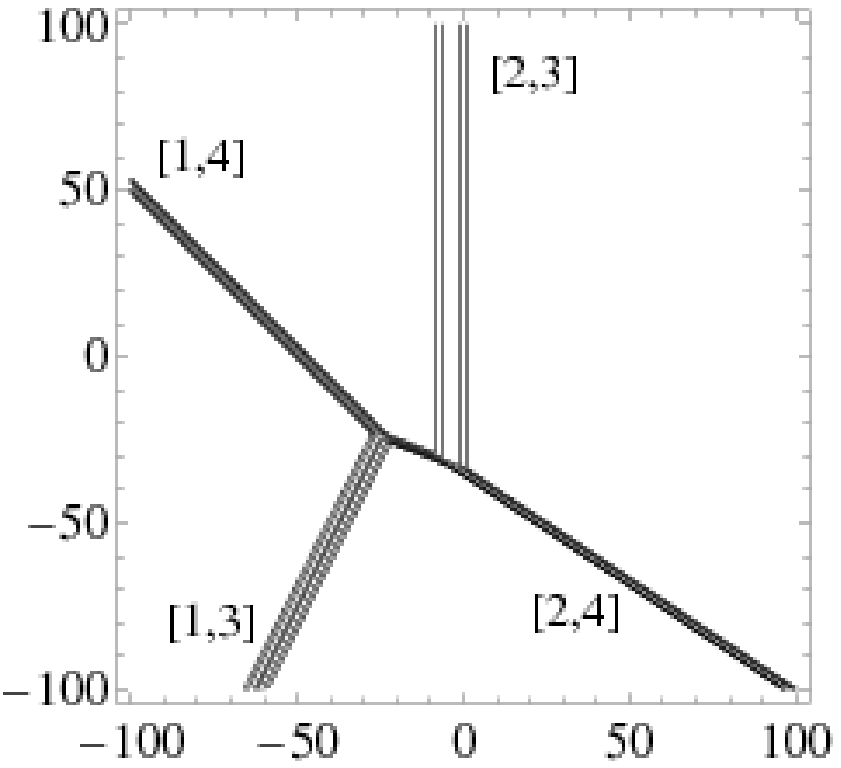}
\caption{Two different $(2,2)$-soliton solutions of KPII with phase
parameters~$(k_1,\dots,k_4)=(-1,-\frac12,\frac12,2)$. The first solution with $\pi=(3421)$
at $t=-16$ and $t=16$ is shown respectively, in (a) \& (b); while (c) \& (d)
represent the second solution with $\pi=(4312)$ at $t=-16$ and $t=16$. Note the space-time
reversal symmetry relating the two solutions in (a), (d) and in (b), (c).}
\label{f:2s2a}
\end{figure}

\noindent {\bf 3 positive parameters}:\, There are two possibilities namely,
$A_2$ with non-zero parameters $a,\,b,\, d$; and $A_1$ with 3 free parameters.
Consider $A_1$ first. In this case neither
$b$ nor $c$ can vanish in $A_1$, else $A_1(34) < 0$. Hence, a 3-parameter 
family of the matrix $A_1$ with non-negative maximal minors
arises in three possible ways namely, $a=0$, or $d=0$, or $A_1(34)=0$
but neither of $a,b,c,d$ is zero. That is, 
\begin{equation}
(i)\,\, A_1= \begin{pmatrix}
 1 &0 &0 &\!-b \\
 0 &1 &c &d \\
\end{pmatrix}, \quad 
(ii)\,\, A_1= \begin{pmatrix}
 1 &0 &\!-a &\!-b \\
 0 &1 &c &0 \\
\end{pmatrix}, \quad 
(iii)\,\, A_1= \begin{pmatrix}
 1 &0 &\!-a &\!-b \\
 0 &1 &c &d \\
\end{pmatrix} \, {\rm with}\,~ A_1(34)=0\,.
\label{e:A1}
\end{equation}
Case $(i)$ in Eq.~\eqref{e:A1} gives rise to a $(2,2)$-soliton
solution with asymptotic line-soliton pairs $([2,4],[1,3])$ as $y\to\infty$,
and $([2,3],[1,4])$ as $y\to -\infty$. The associated pairing map is given by
$\pi_1 =( 3421)$. The second matrix in Eq.~\eqref{e:A1} 
corresponds to another $(2,2)$-soliton solution whose asymptotic line-soliton
pairs are $([1,4], [2,3])$ as $y\to \infty$, and $([1,3],[2,4])$ as $y\to -\infty$.
These two solutions are related via space-time reversal $(x,y,t) \to (-x,-y,-t)$
as shown in Figure~\ref{f:2s2a}. Furthermore, the pairing map for the second solution 
$\pi_2 = (4312)$ satisfies $\pi_2=\pi_1^{-1}$.
The last case in Eq.~\eqref{e:A1} yields a $(2,2)$-soliton with asymptotic line 
soliton pairs $([2,4],[1,2])$ as $y\to\infty$, and $([1,3],[3,4])$ as $y\to -\infty$,
which determine the pairing map $\pi_3=(2413)$. Like the previous two cases,
case $(iii)$ is also related via a space-time reversal to another 
$(2,2)$-soliton solution which is associated with the coefficient matrix 
$A_2$ with 3 positive parameters as in Eq.~\eqref{e:A2sol}. Thus, $A_2$
corresponds to the asymptotic line-soliton pairs $([1,3],[3,4])$ 
as $y\to\infty$ and $([2,4],[1,2])$ as $y\to -\infty$, with the pairing map 
$\pi_4=(3142)=\pi_3^{-1}$.

\noindent {\bf 2 positive parameters}:\, It should be clear from Eq.~\eqref{e:A2sol}
that if either $b=0$ or $c=0$ in $A_1$, it can not satisfy Condition \ref{positive}.
So, $A_1$ will contain precisely 2 free parameters if and only if $a=d=0$. 
Similarly, $A_2$ in  Eq.~\eqref{e:A2sol} will have 2 free parameters if 
and only if $b=0$. Each of the nonzero parameters $b,c$ in $A_1$, and
$a,d$ in $A_2$ can be rescaled to unity without loss of generality, 
by utilizing the gauge freedom described in Property \ref{p:tau}(v) 
in Section \ref{s:general}.
The matrices resulting in this way from $A_1$ and $A_2$ are denoted by
$A_{\mathrm{P}}$ and $A_{\mathrm{O}}$ respectively, and are presented
below in Eq.~\eqref{e:Acanonical}. Each coefficient matrix produces an 
equivalence class
of $2$-soliton solutions with asymptotic soliton pairs $([1,4],[2,3])$ for 
$A_{\mathrm{P}}$, and $([1,2],[3,4])$ for $A_{\mathrm{O}}$, as $|y|\to \infty$.

\subsection{Equivalence classes of 2-soliton solutions}
As noted in the previous subsection, there are three types of 2-soliton solutions.
They will be referred to as O-, T- and P- types (following the
terminology introduced in Ref.~\cite{Kodama}) below. These are
identified by the canonical coefficient matrices
\begin{gather}
A_{\mathrm{O}}=
\begin{pmatrix}1 &1 &0 &0\\ 0 &0 &1 &1\end{pmatrix},
\qquad
A_{\mathrm{T}}=
\begin{pmatrix}1 &0 &-1 &-1\\ 0 &1 &x_1 &x_2\end{pmatrix},
\qquad
A_{\mathrm{P}}=
\begin{pmatrix}1 &0 &0 &-1\\ 0 &1 &1 &0\end{pmatrix},
\label{e:Acanonical}
\end{gather}
with $x_1>x_2>0$ in~$A_{\mathrm{T}}$.
Up to rescaling of columns, the above matrices were also obtained 
as special subcases of the classification presented in Section~\ref{s:twosolitons}.1.
$A_{\mathrm{O}}$ describes O-type 2-solitons,
with asymptotic line-solitons [1,2] and [3,4];
$A_{\mathrm{T}}$ describes T-type resonant 2-solitons
with asymptotic line-solitons [1,3] and [2,4];
and $A_{\mathrm{P}}$ describes P-type 2-solitons
with asymptotic line-solitons [1,4] and [2,3].
A distinctive feature of the pairing map $\pi$ of a
$2$-soliton solution is the fact that $\pi$ is an involution
satisfying $\pi= \pi^{-1}$. As a result, $\pi$ can be expressed as
a product of disjoint $2$-cycles (see e.g.,~\cite{Bona}). For the 
permutation group ${\mathcal{S}}_4$, there are
only 3 such involutions corresponding to the total number of disjoint partitions
of $[4]$ into 2 pairs. In cycle notation, these involutions are given by
$\pi_{\mathrm{O}} = (12)(34),\,\pi_{\mathrm{T}}=(13)(24)$ and $\pi_{\mathrm{P}}=(14)(23)$
for the O-, T- and P-type $2$-soliton solutions, respectively. The $2$-cycles are also
evident from the chord diagrams illustrated by Figure~\ref{f:chord2}
of the previous section where the extreme left (3412) 
diagram corresponds to the T-type $2$-soliton, while the extreme right diagrams (4321) 
and (2143) respectively, represent the P- and O-type $2$-soliton equivalence classes.

\begin{figure}[t!]
\centering
\raisebox{0.83in}{(a)}\includegraphics[scale=0.55]{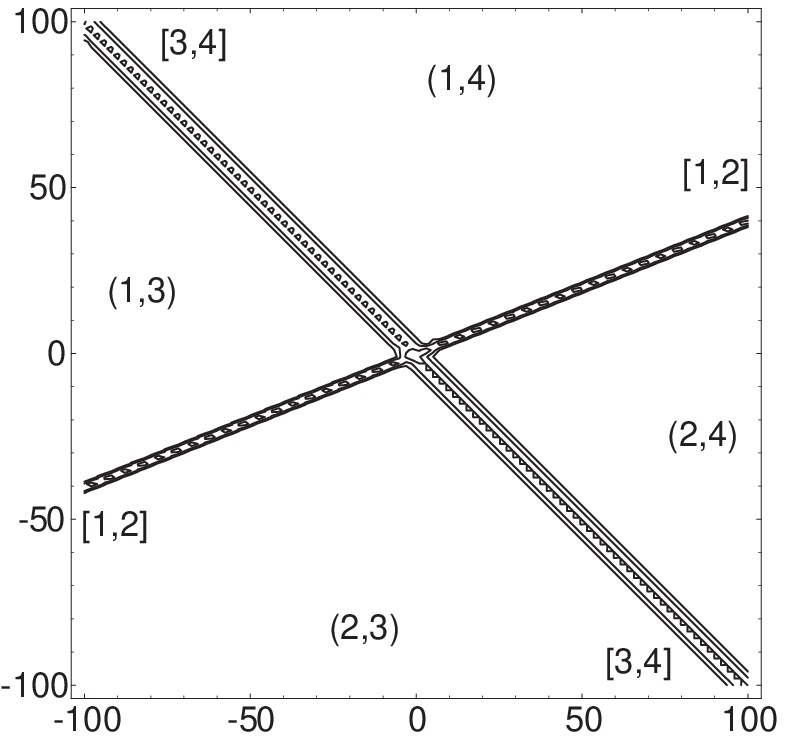} \hskip 0.4cm
\raisebox{0.83in}{(b)}\raisebox{-0.1in}{\includegraphics[scale=0.55]{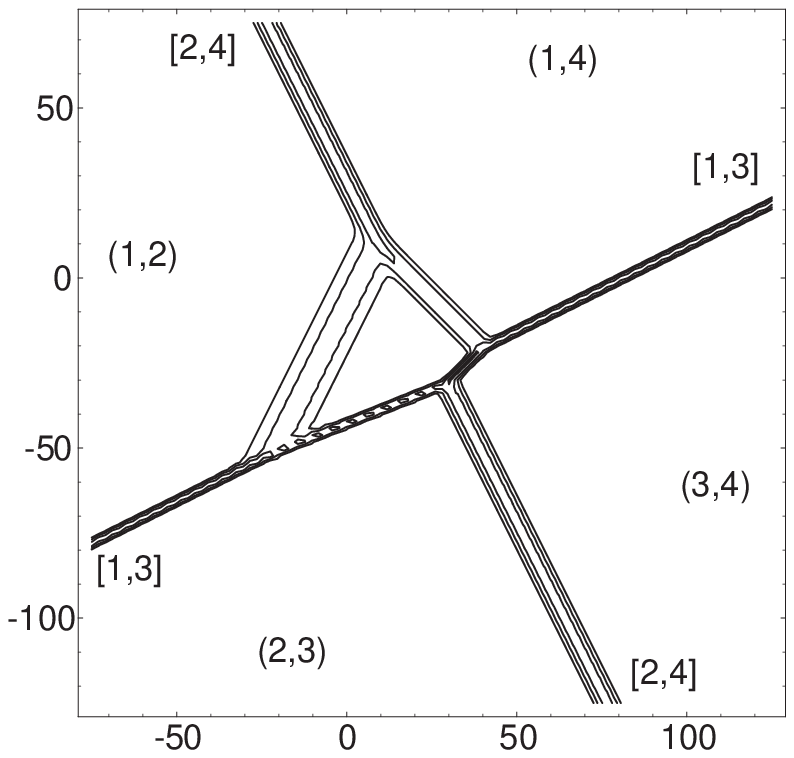}} \hskip 0.5cm
\raisebox{0.83in}{(c)}\includegraphics[scale=0.55]{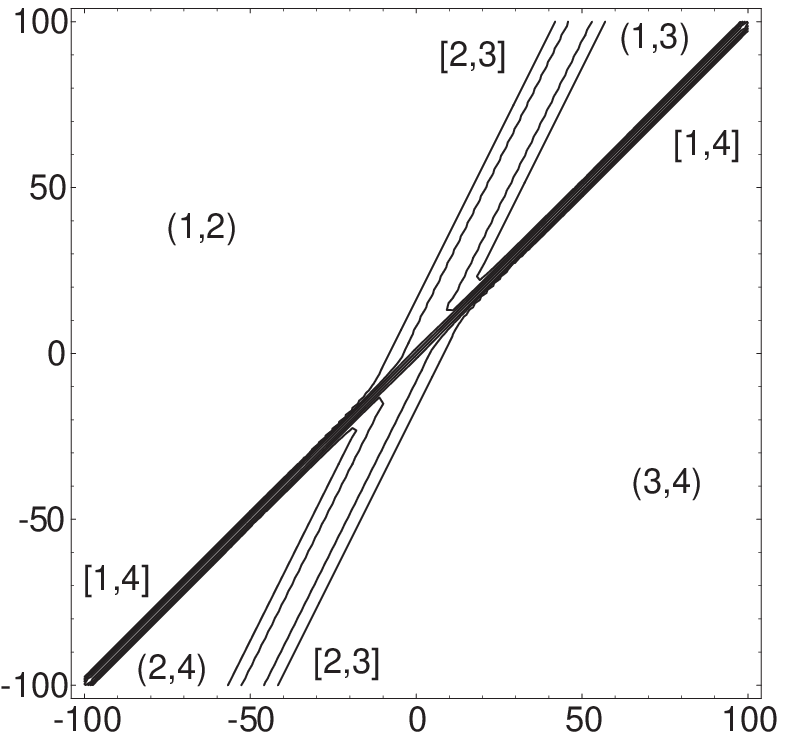} 
\caption{Three different two-soliton solutions of KPII with the same
phase parameters~$(k_1,\dots,k_4)=(-2,-\frac12,0,1)$, 
illustrating the three equivalence classes: 
(a)~O-type 2-soliton solution, yielding $(c_1,c_2)=(-\frac52,1)$ and $(a_1,a_2)=(\frac32,1)$;
(b)~T-type 2-soliton solution, yielding $(c_1,c_2)=(-2,\frac12)$ and $(a_1,a_2)=(2,\frac32)$;
(c)~P-type 2-soliton solution, yielding $(c_1,c_2)=(-1,-\frac12)$ and
$(a_1,a_2)=(3,\frac12)$.
(Note that for this P-type solution $c_{[1,4]}<c_{[2,3]}$.)}
\label{f:2solitonphases}
\end{figure}

Figure~\ref{f:2solitonphases} shows a representative solution
for each of the three equivalence classes with the same phase 
parameters $k_1,\dots,k_4$.
Note that the O- and P-type solitons interact via an X-junction (ignoring the phase shifts), while 
the T-type solitons interact via four
Y-junctions connecting the four asymptotic line-solitons to four
intermediate segments. Each of these intermediate segments satisfy the
nonlinear dispersion relation~\eqref{e:dispersionrelation}
and at each Y-junction the resonance condition~\eqref{e:resonance}
is satisfied. Thus each intermediate segment is also a line-soliton.
For example, in Figure~\ref{f:2solitonphases}(b) the asymptotic line 
soliton $[1,3]$ (as $y \to -\infty$) forms the intermediate line-solitons
$[1,2]$ and $[2,3]$ at the bottom left Y-junction. The line-soliton $[2,3]$ 
connects with the asymptotic line-soliton $[2,4]$ (as $y \to \infty$)
and the line-soliton $[1,2]$ connects with the asymptotic line 
soliton $[2,4]$ (as $y \to -\infty$). Similarly, the asymptotic line
soliton $[1,3]$ (as $y \to \infty$) forms the intermediate line-solitons
$[1,4]$ and $[3,4]$ at the top right Y-junction. The line-soliton $[3,4]$
connects with the asymptotic line-soliton $[2,4]$ (as $y \to \infty$)
and the line-soliton $[1,4]$ connects with the asymptotic line
soliton $[2,4]$ (as $y \to -\infty$). 

An important distinction among the three types of 2-soliton solutions
is that they cover different regions of the soliton parameter space.
Suppose $(a_1,c_1)$ and $(a_2,c_2)$ are the soliton parameters of the
asymptotic line-solitons of each type with the same set of distinct phase 
parameters. Since the phase parameters are ordered:~$k_1<\dots<k_4$,
the soliton parameters satisfy the following relations 
which can be easily verified using Eqs.~\eqref{e:solitonparameters}.
\begin{enumerate}
\advance\itemsep-4pt
\item
For an O-type 2-soliton solution $c_2>c_1$ and $c_2-c_1>a_1+a_2$.
\item 
For a T-type 2-soliton solution $c_2>c_1$, and $|a_1-a_2|<c_2-c_1<a_1+a_2$.
\item
For a P-type 2-soliton solution $a_2>a_1$ and  $|c_2-c_1|<a_2-a_1$.
\item
$(c_2-c_1)_\mathrm{O}>(c_2-c_1)_\mathrm{T}>|c_2-c_1|_\mathrm{P}$\,, \quad
$(a_1+a_2)_\mathrm{O}<(a_1+a_2)_\mathrm{T}=(a_1+a_2)_\mathrm{P}$\,, and \\
$|a_2-a_1|_\mathrm{O}=|a_2-a_1|_\mathrm{T}<(a_2-a_1)_\mathrm{P}$\,.
\end{enumerate}
Note that for O- and T-type solutions the soliton directions are ordered
while for P-type solutions the amplitudes are ordered. {\em Any} choice
of the soliton parameters $\{a_i\,, c_i\,|a_i > 0\}_{i=1}^2$ would lead to one
of the three types of 2-soliton solutions provided that 
$\{c_1 \pm a_1,\, c_2 \pm a_2\}$ are distinct real numbers.
Thus, the three types of 2-soliton solutions divide the soliton parameter
space into disjoint sectors bounded by the hyperplanes
$|c_2-c_1|= a_1 + a_2$ and $|c_2-c_1|= |a_1 - a_2|$.
At each boundary between two disjoint regions of the soliton
parameter space two of the phase parameters coincide.
In such situation, it can be shown (by taking suitable limits)
that the 2-soliton solution degenerates into a 
Y-junction~\cite{jphysa36p10519,Kodama}.

Yet another difference is in the phase shifts experienced by the asymptotic
line-solitons of each type. The position of an asymptotic line-soliton $[i,j]$
is determined by the dominant phase combinations across the
soliton and from the asymptotic formulas Eqs.~\eqref{e:tauasymp}
and \eqref{e:uasymp}. The pairs of dominant phase combinations across 
the soliton $[i,j]$ as $y\to \infty$ and as $y\to -\infty$ are {\em distinct} 
from each other (see e.g., Figure~\ref{f:2solitonphases}). This
fact gives rise to the phase (position) shift $\Delta_{ij}=\delta^+_{ij}-\delta^-_{ij}$
for the asymptotic line-soliton $[i,j]$. Since the asymptotic positions 
$\delta_{ij}^{\pm}$ have the {\em same} linear dependence on time $t$, the 
phase shift $\Delta_{ij}$ is independent of $t$.
The dominant phase combinations across
an asymptotic line-soliton in each of the three types of 2-soliton solutions
can be determined from the asymptotic analysis of section \ref{s:general} and
are shown in Figure~\ref{f:2solitonphases}. Then the soliton phase shifts  
computed using the parameters $(a_1,c_1),\,(a_2,c_2)$ and the canonical coefficient 
matrices in Eq.~\eqref{e:Acanonical} are given by
\begin{equation}
\Delta_\mathrm{O}= 
\log\frac{(c_1-c_2)^2-(a_1-a_2)^2}{(c_1-c_2)^2-(a_1+a_2)^2}
= \Delta_\mathrm{P}\,, \qquad
\Delta_\mathrm{T}= \log\frac {(c_1-c_2)^2-(a_1-a_2)^2}{(c_1-c_2)^2-(a_1+a_2)^2}
 + \log\left(\frac{x_1}{x_2}-1\right)\,,
\label{e:phaseshift}
\end{equation}
The phase shifts experienced by the pair of asymptotic line-solitons are 
of opposite signs in each of the three cases. In above, we assume
that $c_1 \neq c_2$ (i.e., the solitons are not parallel), then
$\Delta$ denotes the phase shift of the soliton with direction parameter 
$c_2$ where $c_2>c_1$. For non-resonant (O- and P-type) 2-soliton solutions,
the phase shift has only one and the same term which depends symmetrically on 
the soliton parameters. However it is easy to verify from the various
inequalities mentioned above, among the soliton parameters 
that $\Delta_{\mathrm{O}}$ is always {\em positive} while $\Delta_{\mathrm{P}}$ 
is always {\em negative}.
For resonant 2-soliton solutions there is an additional term 
that depends on the free parameters $x_1, \, x_2$ of the coefficient matrix $A_{\mathrm{T}}$
in Eq.~\eqref{e:Acanonical}. In this case, the phase shift $\Delta_{\mathrm{T}}$ is not
sign-definite unlike the other two cases.
\section{Duality and $N$-soliton solutions}
\label{s:elastic}
In the previous section, we described two types of equivalence classes 
of line-soliton solutions of KPII: (i)\, the $(2,2)$-solitons 
where the sets $S_{\pm}$ (cf. Definition~\ref{D:equiv})
of asymptotic line-solitons
as $y \to \pm \infty$ are distinct, and (ii)\, the $2$-soliton solutions
with $S_-=S_+$. Furthermore, we noted that there are pairs of distinct equivalence
classes of $(2,2)$-soliton solutions related via space-time reversal
(cf. Figure~\ref{f:2s2a}). The general $(N_-,N_+)$-soliton 
solutions can be also categorized in a similar fashion
according to whether the sets of asymptotic line-solitons $S_+$ and 
$S_-$ are distinct, or if $S_-=S_+$. In the first case, pairs of equivalence
classes of solutions are related by space-time reversal, while the latter corresponds 
to the special case of $N$-soliton solutions to be discussed in
this section.
\subsection{Duality of line-solitons}
The KPII equation~\eqref{e:KP} is invariant under the inversion
$(x,y,t) \to (-x,-y,-t)$. Consequently, if $u(x,y,t)$ is a 
$(N_-,N_+)$-soliton solution of KPII with given sets $S_\pm$ of
asymptotic line-solitons and with $N_-=M-N$ and $N_+=N$, 
then $u(-x,-y,-t)$ is a $(N,M-N)$-soliton solution with {\em reversed}
sets $S_{\mp}$ of asymptotic line-solitons. We refer to the solutions 
$u(x,y,t)$ and $u(-x,-y,-t)$, 
as well as their respective equivalence classes as {\em dual} to each other.
Let $\tau_{NM}(x,y,t)$ denote the $\tau$-function in Eq.~\eqref{e:tauexp},
generating the solution $u(x,y,t)$ via Eq.~\eqref{e:u}, then 
the solution $u(-x,-y,-t)$ will be generated by $\tau_{NM}(-x,-y,-t)$ 
as Eq.~\eqref{e:u} remains invariant under $(x,y,t) \to (-x,-y,-t)$. 
Note however that $\tau_{N,M}(-x,-y,-t)$ 
does not have the form given by Eq.~\eqref{e:tauexp}, but it is indeed possible to construct 
a certain $\tau$-function $\tau_{M-N,M}(x,y,t)$ from $\tau_{N,M}(-x,-y,-t)$ that is
dual to $\tau_{N,M}(x,y,t)$. We describe below how
to construct the function $\tau_{M-N,M}$ from $\tau_{N,M}$.

First we obtain a coefficient matrix for the $\tau$-function $\tau_{M-N,M}$
from the $N \times M$ matrix $A$ associated with $\tau_{N,M}(x,y,t)$.
Since $A$ is of full rank, its rows form a basis for
a $N$-dimensional subspace $W$ of $\Real^{M}$. Let $W^{\perp}$ be the
orthogonal complement of $W$ with respect to the standard inner product
on $\Real^{M}$, and let  $A'$ be a $(M-N)\times M$ matrix whose rows form 
a basis for $W^{\perp}$. Clearly, $A'$ is not unique, but a particular
choice for $A'$ is as follows: Suppose the pivot and non-pivot columns 
of $A$ in RREF are represented by the identity
matrix $I_N$ and the $N\times(M-N)$~matrix $G$ respectively, then  
\begin{equation}
A=[I_N,G]\,P \qquad \Rightarrow \qquad A'=[-G^T, I_{M-N}]\,P\,,
\label{e:A'}
\end{equation}
where $G^T$ is the matrix transpose of $G$, and $P$ is a $M \times M$ 
permutation matrix satisfying $P^T = P^{-1}$. It can be directly verified
from Eq.~\eqref{e:A'} that $AA'^T = 0$, which constitute the orthogonality
relations among the row vectors of $A$ and $A'$. 
Moreover, it was shown in Ref.~\cite{BC} that
the $N \times N$ minors of $A$ and the $(M-N) \times (M-N)$ minors of the matrix
\begin{equation*}
B' := A'E \,, \qquad E:= \diag(-1,1,-1,\ldots,\pm 1)
\end{equation*}
are related as  
\begin{equation*}
A(m_1,\dots,m_N)
   = (-1)^\sigma\,\det(P)\,\,B'(l_1,\dots,l_{M-N})\,,
\end{equation*} 
where $\sigma= M(M+1)/2+N(N+1)/2$, and where 
$\{m_1,m_2,\ldots,m_N\}$ is the complement of $\{l_1,l_2,\ldots,l_{M-N}\}$ in $[M]$.
The factor $(-1)^\sigma\,\det(P)=\pm 1$ which depends only on $A$, can be
rescaled by an orthogonal transformation $B' \to OB'\,, \quad \det(O)=\pm 1$, so
that the maximal minors of the rescaled matrix   
\begin{subequations}
\begin{equation}
 B := OB' = O[-G^T, I_{M-N}]\,PE \,, 
\label{e:b}
\end{equation}
satisfies the precise complementarity conditions 
\begin{equation}
A(m_1,\dots,m_N) = B(l_1,\dots,l_{M-N})\,.
\label{e:bminors2}
\end{equation}
\end{subequations}
Hence, if all maximal minors of $A$ are non-negative, then the same 
holds for $B$. 

The $(M-N) \times M$ matrix $B$ plays the role of a coefficient 
matrix for the $\tau$-function $\tau_{M-N,M}(x,y,t)$ which is 
related to the function $\tau_{N,M}(-x,-y,-t)$.
Indeed, if we initially set $\theta_{m,0}=0, \,\, \forall\, m=1,\ldots,M$                         
in Eq.~\eqref{e:tauexp} using Property \ref{p:tau}(v),
then under the transformation $(x,y,t) \to (-x,-y,-t)$, Eq.~\eqref{e:tauexp} yields,  
\begin{equation*}
\tau_{N,M}(-x,-y,-t)= \exp[-\theta(1,\dots,M)]\, \tau'(x,y,t)\,.
\end{equation*}
Using Eq.~\eqref{e:bminors2} and taking the sum over the complementary
indices $l_1,\ldots,l_{M-N}$ (instead of $m_1,\ldots,m_N$), the function 
$\tau'(x,y,t)$ can be expressed as
\begin{equation}
\tau'(x,y,t)=  \hspace{-0.3 in}
  \sum\limits_{1\le l_1<\dots<l_{M-N}\le M}\hspace{-0.3 in}
    V(m_1,\dots,m_N)\,\,
    B(l_1,\dots,l_{M-N})\,\,
    \exp[\theta(l_1,\dots,l_{M-N})]\,,
\label{e:tauprime}
\end{equation}
where 
\begin{equation*}
V(m_1,\dots,m_N) = \prod_{1\le s < r\le N}(k_{m_{r}}-k_{m_{s}})\,
\end{equation*}
are the Van~der~Monde coefficients as in Eq.~\eqref{e:tauexp}. 
Next, if we introduce new phase constants by
\[ \theta_{m,0} = \sum_{r \neq m}\ln|k_T-k_m|\,, \qquad m=1,\ldots,M \]
which satisfy the identity
\[\exp[\theta_{l_1,0}+\ldots+\theta_{l_{M-N},0}] = 
\frac{V(l_1,\dots,l_{M-N})V(1,2,\ldots,M)}{V(m_1,\dots,m_N)}\,,\]
and make the replacement:\, $\theta_m \to \theta_m+\theta_{m,0}$ in 
Eq.~\eqref{e:tauprime}, we finally obtain the dual $\tau$-function
\begin{equation}
\tau_{M-N,M}(x,y,t):= \frac{\tau'(x,y,t)}{V(1,\ldots,M)} = \hspace{-0.1 in} 
  \sum\limits_{1\le l_1<\dots<l_{M-N}\le M}\hspace{-0.3 in}
    V(l_1,\dots,l_{M-N})\,\,
    B(l_1,\dots,l_{M-N})\,\,
    \exp[\theta(l_1,\dots,l_{M-N})]\,.                          
\label{e:taudual}
\end{equation}
It is clear from Eqs.~\eqref{e:u} and ~\eqref{e:taudual} 
that the functions $\tau_{N,M}(-x,-y,-t),\, \tau'(x,y,t)$, 
and $\tau_{M-N,M}(x,y,t)$ give rise to the same solution $u(-x,-y,-t)$ of KPII.
Thus we have the following.
\begin{proposition}
(i)\,\, The equivalence classes of solutions generated by the $N\times M$
coefficient matrix $A$ and the $(M-N) \times M$ matrix $B$ defined by 
Eq.~\eqref{e:b}, are dual to each other. If a $(M-N,N)$-soliton solution
$u(x,y,t)$ of KPII belongs to a certain equivalence class, then its dual equivalence class
contains the $(N,M-N)$-soliton solution $u(-x,-y,-t)$. \\
(ii)\,\, Let $\{m_1,\dots,m_N\}$ and $\{l_1,\dots,l_{M-N}\}$ be a disjoint partition
of the integer set $[M]$, then $\theta(m_1,\ldots,m_N)$ is a phase combination
present in the $\tau$-function $\tau_{N,M}$ if and only if $\theta(l_1,\dots,l_{M-N})$ is 
a phase combination in the dual $\tau$-function $\tau_{M-N,M}$. \\
(iii)\,\, If $\pi \in \mathcal{S}_M$ is the pairing map for a given equivalence class, then the
pairing map for the dual equivalence class is given by $\pi^{-1}$.
\label{P:duality}
\end{proposition}
Proposition \ref{P:duality} establishes a one-to-one correspondence between an 
equivalence class and its dual. Indeed from Eq.~\eqref{e:symm} in 
Section~\ref{s:general}.3, we have the following result for the dual equivalent
classes.
\begin{proposition}
For given positive integers $M, N$, with $N<M$, the number of of distinct 
equivalence classes of the $(M-N,N)$-soliton
solutions are exactly the same as the number of dual equivalence classes
of $(N,M-N)$-soliton solutions.
\label{P:symmetry}
\end{proposition}
\begin{remark}
The open chord diagrams of the pairing maps $\pi$ and $\pi^{-1}$ which
correspond to a line-soliton equivalence class and its dual are related to each
other via a reflection about the horizontal line together with reversing the 
direction of the chords. This is due to the fact that the exedance set of 
$\pi^{-1}$ is given by $\{\pi(g_n)\}_{n=1}^{M-N}$, while the anti-exedance set
is given by  $\{\pi(e_n)\}_{n=1}^{N}$. This transformation on the chord diagrams
can be regarded as the combinatorial analogue of the inversion symmetry 
$(x,y,t) \to (-x,-y,-t)$. acting on the solutions $u(x,y,t)$ of the KPII line
solitons.
\end{remark}
When $M=2N$, it follows from Proposition \ref{P:pairing} that $N_-=N_+ = N$,
which leads to the $(N,N)$-soliton solutions. In this
case the number of asymptotic line-solitons as $y \to \infty$ and
as $y \to -\infty$  are the same, but in general,
the amplitudes and directions of the line-solitons 
will be different as seen for the $(2,2)$-soliton examples in Figure~\ref{f:2s2a} of
Section~\ref{s:twosolitons}. A particularly interesting subclass of the
$(N,N)$-solitons are the $N$-soliton solutions which were introduced in
Section~\ref{s:introduction}, and which
are characterized by identical sets of asymptotic
line-solitons as $|y| \to \infty$, i.e., $S_- = S_+$. We discuss them next.

\subsection{The $N$-soliton solutions}
This special family of line-soliton solutions of KPII consists
of equivalence classes that are invariant
under the inversion symmetry $(x,y,t) \to (-x,-y,-t)$. That is, each
equivalence class is its own dual so that both
$u(x,y,t)$ and $u(-x,-y,-t)$ belong to the same equivalence class of solutions.
This implies that for each $N$-soliton solution $u(x,y,t)$, the asymptotic line 
solitons arise in pairs, where each pair consists of a line 
soliton as $y \to \infty$ and another soliton as $y \to -\infty$, moreover,
both line-solitons have identical amplitude and direction.
Thus, a $(N_-,N_+)$-soliton solution of KPII is a $N$-soliton solution 
if and only if it is {\em self-dual}, i.e., if and only if the index sets labeling 
the asymptotic line-solitons satisfy $S_- = S_+$.
The main features of the $N$-soliton solutions follow from 
from Propositions~\ref{P:pairing}, \ref{P:derangement}, \ref{P:duality} and 
Definition~\ref{D:equiv}. These are listed below.
\begin{property}
\begin{enumerate}
\item
The $\tau$-function of an $N$-soliton solution, denoted by $\tau_N:=\tau_{N,2N}$, is expressed
in terms of $2N$ distinct phase parameters and an $N \times 2N$   
coefficient matrix $A$ which satisfies Condition~\ref{positive}. Then it
follows from Eq.~\eqref{e:bminors2} that the 
$N \times N$ minors of $A$ satisfy the duality conditions:
\begin{equation}
A(m_1,\ldots,m_N) = 0 \qquad \Longleftrightarrow \qquad A(l_1,\ldots,l_N) = 0\,,
\label{e:dualminor}
\end{equation}
where the indices $\{m_1,\ldots,m_N\}$ and $\{l_1,\ldots,l_N\}$ form a disjoint 
partition of integers $\{1,2,\ldots, 2N\}$. That is, the phase combination
$\theta(m_1,\ldots,m_N)$ is present in $\tau_{N}$ if and only if 
$\theta(l_1,\ldots,l_N)$ is.
\item
Each $N$-soliton solution exactly $N$ 
asymptotic line-solitons as $y \to \pm \infty$ identified by the
same index pairs $[e_n,g_n]$ with $e_n < g_n, \, n=1,\ldots, N$.
The sets $\{e_1, \ldots, e_N\}$ and $\{g_1, \ldots, g_N\}$ label
respectively, the pivot and non-pivot columns of the coefficient matrix
$A$. Hence, they form a disjoint partition of the integer set $[2N]$.
\item
The amplitude and direction of the $n^{\mathrm {th}}$ asymptotic line
soliton $[e_n,g_n]$ are the same as $y \to \pm \infty$, and
are given in terms of the phase parameters as $a_n = k_{g_n} - k_{e_n}$
and $c_n = k_{g_n} + k_{e_n}$.
\item
The pairing maps associated with $N$-soliton solutions are involutions
of $\mathcal{S}_{2N}$ with no fixed points, defined by the set
$\mathcal{I}_{2N} = \{\pi \in \mathcal{S}_{2N}|\pi^{-1}=\pi,\, \pi(i) \neq i,\, \forall\, i \in [2N]\}$. 
Such permutations can be expressed as products of $N$ disjoint $2$-cycles,
and theirs chord diagrams are self-dual, i.e. symmetric about the horizontal
axis (see e.g., the 2-soliton chord diagrams in Section \ref{s:twosolitons}. The total 
number of such involutions
is given by $|I_{2N}| = (2N-1)!!= 1\cdot3\ldots\cdot(2N-1)$~\cite{Bona}.
Hence, there are $(2N-1)!!$ distinct equivalence classes of $N$-soliton solutions. 
\end{enumerate}
\label{p:Nsoliton}
\end{property}
Examples of the $N$-soliton solutions with special choices
of the functions $\{f_n\}_{n=1}^N$ in Eq.~\eqref{e:f} and the coefficient matrix
$A$ are given below.
\begin{example}
{\bf O-type $N$-soliton solutions.}
These are the well-known \cite{PLA95p1,MatveevSalle} multi-soliton solutions
of KPII constructed by choosing $\{f_n\}_{n=1}^N$ according to
\begin{equation*}
f_n(x,y,t)= e^{\theta_{2n-1}}+e^{\theta_{2n}}\,, \quad n=1,\dots,N\,.
\label{e:ordinaryNsoliton}
\end{equation*}
The corresponding coefficient matrix is given by
\begin{equation*}
A_{\mathrm{O}} =
  \begin{pmatrix}
     1 & 1 & 0 & 0 & \cdots & 0 & 0 \\
     0 & 0 & 1 & 1 & \cdots & 0 & 0 \\
     \vdots & \vdots & \vdots &\vdots & \vdots & \vdots & \vdots\\
     0 & 0 & 0 & 0 & \cdots & 1 & 1 \\
  \end{pmatrix}\,,
\label{e:Aordinary}
\end{equation*}
with $N$ pairs of identical columns at positions $\{2n-1,2n\},\, n=1,\dots,N$.
Thus, there are $2^N$ nonzero maximal minors of $A_{\mathrm{O}}$,
given by $A_{\mathrm{O}}(m_1,\ldots, m_N) = 1$, where $m_n = 2n-1$ or $m_n=2n$
for $n=1,\ldots,N$. The $N$ asymptotic line-solitons are identified
by the index pairs $\{[2n-1,2n]\}_{n=1}^N$, and the corresponding permutation is
$\pi=(2,1,4,3,\ldots,2N,2N-1)$ (or $\pi=(12)(34)\cdots(2N,2N-1)$ in the cycle notation).
The amplitude and direction of the
$n$-th soliton are $a_n= k_{2n}-k_{2n-1}$ and $c_n= k_{2n-1}+k_{2n}$, respectively.
Note that the soliton directions are ordered as $c_1 < c_2 < \ldots < c_N$
due to the ordering of the phase parameters $k_n$.
In fact, the soliton parameters satisfy the inequalities:\,
$ c_{n+1} - c_n > a_n + a_{n+1},\quad n=1,2,\ldots,N-1$.
Therefore the asymptotic line-solitons for the O-type solutions can not take
arbitrary values of amplitude and direction, and thus do not cover the entire
soliton parameter space as was already noted for the $2$-soliton case in
Section~\ref{s:twosolitons}.2.
Apart from the position shift of each soliton,
the interaction gives rise to a pattern of $N$ intersecting
lines in the $(x,y)$-plane, as shown in Figure~\ref{f:3s}(a).
%
\begin{figure}[t!]
\raisebox{0.85in}{(a)}\includegraphics[scale=0.55]{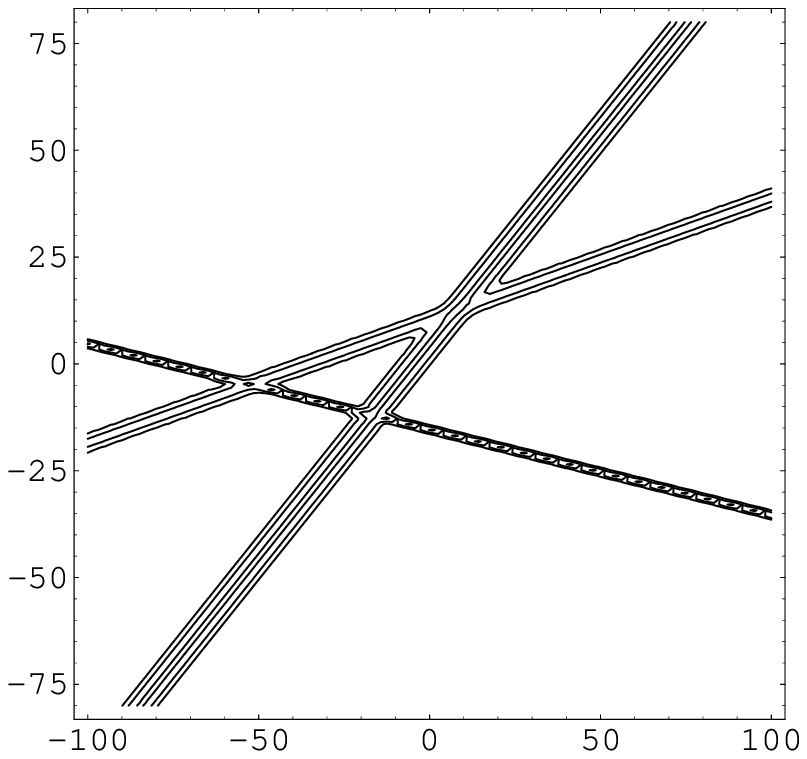} \hskip 0.4cm
\raisebox{0.85in}{(b)}\includegraphics[scale=0.55]{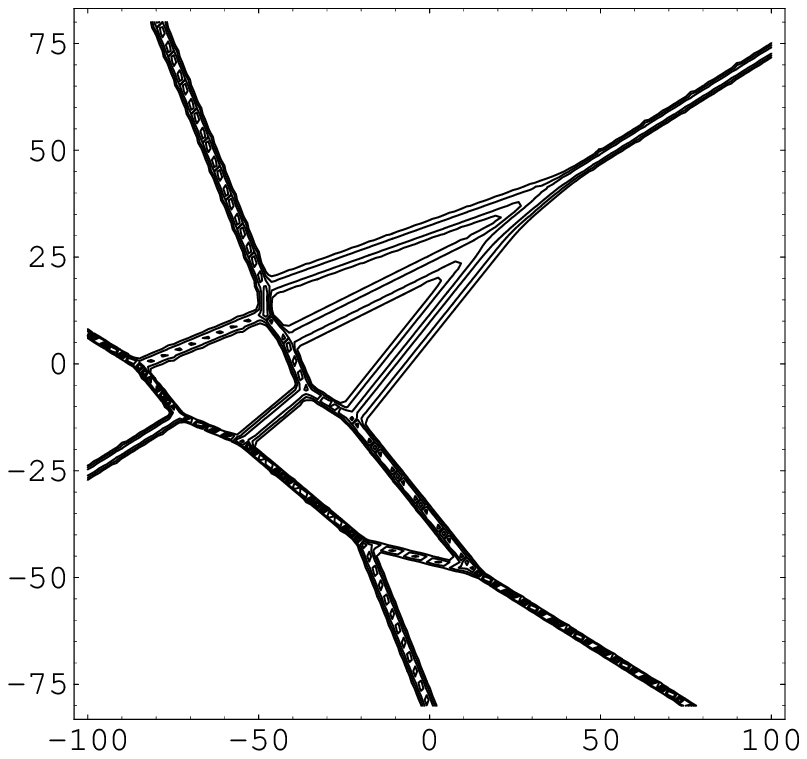} \hskip 0.5cm
\raisebox{0.85in}{(c)}\includegraphics[scale=0.55]{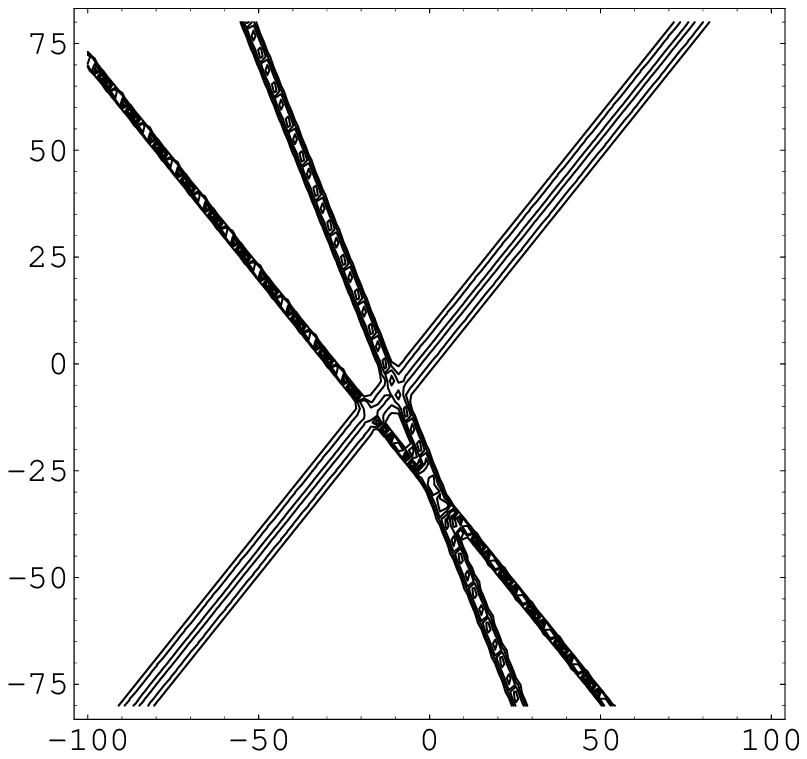} 
\caption{Three different $3$-soliton solutions of KPII with the same
phase parameters~$(k_1,\dots,k_6)=(-3,-2,0,1,-\frac32,2)$,
illustrating the three equivalence classes:
(a)~O-type, (b)~T-type, and (c)~P-type 3-soliton solutions.}
\label{f:3s}
\end{figure}

\medskip

\noindent {\bf T-type $N$-soliton solutions.}
These solutions are obtained by choosing the functions 
in Eq.~\eqref{e:f} as
\begin{equation*}
\label{e:resonantNsoliton}
f_n = f^{(n-1)}\,, \quad n=1,\ldots,N \qquad \mbox{with} \quad
f(x,y,t)=\sum_{m=1}^{2N} e^{\theta_m}\,,
\end{equation*}
which yields the coefficient matrix
\begin{equation*}
A_{\mathrm{T}} =
  \begin{pmatrix}
     1 & 1 & \cdots & 1  \\
     k_1 & k_2 & \cdots & k_{2N} \\
     \vdots & \vdots & \ddots & \vdots \\
     k_1^{N-1} & k_2^{N-1} & \cdots & k_{2N}^{N-1}
  \end{pmatrix}\,
\end{equation*}
In this case all of the $N \times N$
minors of the coefficient matrix~$A$ are positive, each
being equal to a Van~der~Monde determinant.
These solutions were investigated in Ref.~\cite{jphysa36p10519} where
it was shown that they also satisfy the finite Toda lattice hierarchy.
The $n^{\mathrm th}$ line-soliton is labeled by the index pair~$[n,n+N]$;
that is, its amplitude and direction are determined by the phase parameters 
$(k_n, k_{n+N})$. The corresponding permutation is $\pi=(N+1,N+2,\ldots,2N,1,2,\dots,N)$
(or $\pi=(1,N+1)(2,N+2)\cdots(N,2N)$)
whose open chord diagram has the maximum number of crossings $N(N-1)/2$.
This implies that the T-type $N$-soliton solution belongs to the top cell of $Gr^+(N,2N)$.
Like the O-type $N$-soliton solutions,
these T-type $N$-soliton solutions do not cover the whole soliton parameter space.
In this case the soliton parameters satisfy the constraints:\,
$|a_{n+1}-a_n| < c_{n+1}-c_n < a_{n+1}+a_n, \quad n=1,2, \ldots, N-1$.
It was also shown in Ref.~\cite{jphysa36p10519}
that these soliton solutions display phenomena of soliton resonance
and web structure as shown in Figure~\ref{f:3s}(b).
Moreover, the intermediate interaction segments are also
line-solitons because they satisfy the dispersion relation~\eqref{e:dispersionrelation}.
All of the asymptotic and intermediate line-solitons interact
via three-wave resonances. That is, at each interaction vertex or Y-junction 
(cf. Figure~\ref{f:kpfig}(c), the three interacting line-solitons satisfy  
Miles' resonance condition given by Eq.~\eqref{e:resonance}.

\medskip

\noindent {\bf P-type $N$-soliton solutions.}
Yet another type of $N$-soliton solutions is obtained by prescribing
\begin{equation*}
f_n(x,y,t)= e^{\theta_n} + (-1)^{N-n}e^{\theta_{2N-n+1}}\,,
\quad n=1,2, \ldots N \,,
\label{e:nonresonantNsoliton}
\end{equation*}
in Eq.~\eqref{e:f}. The coefficient matrix is given by
\begin{equation*}
A_{\mathrm{P}} =
  \begin{pmatrix}
     1 & 0 & \cdots & \cdots & 0 & 0 & \cdots & \cdots & 0 & * \\
     0 & 1 & 0 & \cdots & 0 & 0 & \cdots & 0 & * & 0 \\
 \vdots & \ddots & \ddots & \ddots & \vdots & \vdots & \ddots & \iddots &
 \iddots & \vdots \\
     0 & \cdots & 0 & 1 & 0 & 0 & -1 & 0 & \cdots & 0 \\
     0 & \cdots & 0 & 0 & 1 & 1 & 0 & 0 &\cdots & 0
  \end{pmatrix}\,,
\label{e:Aasymmetric}
\end{equation*}
where the asterisk in the
$(2N-n+1)^{\mathrm{th}}$ column is equal to $(-1)^{N-n}$.
Like $A_{\mathrm{O}}$, the matrix $A_{\mathrm{P}}$ also has $N$ pairs of parallel
columns labeled by $\{(n, \, 2N-n+1),\, n=1,\ldots,N\}$ and
$2^N$ non-vanishing minors, and each nonzero minor is~1.
The $n^\mathrm{th}$ soliton is identified by the index pair $[n,2N-n+1]$, and
the corresponding permutation is $\pi=(2N,2N-1,\ldots,2,1)$ 
(or $\pi=(1,2N)(2,2N-1)\cdots(N,N+1)$).
The soliton direction and amplitude are given by
$c_n = k_n + k_{2N-n+1}$ and $a_n = k_{2N-n+1}-k_n$, respectively.
Note that the soliton directions are not ordered
as in the previous two cases.
In fact, taking $c_1 = c_2 = \ldots = c_N = 0$ yields the reduction to
solutions of the KdV equation \cite{Kodama}.
But the soliton amplitudes
in this case are ordered as $a_1 > a_2 > \ldots > a_N$.
Moreover, the soliton parameters satisfy the constraints:\,
$ |c_{n+1}-c_n| < a_n -a_{n+1},\quad n=1,2, \ldots, N-1$.
These solutions interact non-resonantly, like the O-type $N$-solitons,
i.e., pairwise with an overall phase shift after collision (see Figure~\ref{f:3s}(c)).
However, as in the case of the 2-soliton solutions in section~\ref{s:twosolitons}.2,
the pairwise phase shifts for P-type solitons is of opposite sign from that of
O-type soliton solutions.
\label{E:nsol}
\end{example}
\subsection{Soliton parameters and pairing map}
The most notable differences between the O-, T- and P-type
$N$-soliton solutions in Example \ref{E:nsol}
are that they span different regions of the
soliton parameter space and that they exhibit dissimilar interaction patterns
and phase shifts. However, in addition to these non-resonant and fully resonant 
solutions, a large family of partially resonant solutions exists 
when $N>2$. Thus, the family of $N$-soliton solutions of KPII is much larger 
than previously thought, and their classification is indeed nontrivial. Even for $N=3$, 
Property \ref{p:Nsoliton}(iv) implies that there are 15 inequivalent types of 
$3$-soliton solutions. For increasing values of $N$, it turns out to be a difficult 
task to classify these solutions according to their coefficient matrices,
as was done for $N=2$ in Section~\ref{s:twosolitons}. Instead, a more direct 
approach is to enumerate the $N$-soliton solutions via the involutions
$\mathcal{I}_{2N} \subset \mathcal{S}_{2N}$, by constructing a representative coefficient matrix $A$ for
each $N$-soliton equivalence class starting from a pairing map $\pi \in \mathcal{I}_{2N}$.
In what follows, we describe a slightly modified approach. We start with
the set of amplitudes and directions of the asymptotic line-solitons 
as $|y| \to \infty$. We first recover the soliton pairings from
this physical data, then construct the coefficient matrix $A$ from
the obtained pairing map. This provides a method to algebraically reconstruct 
the $N$-soliton solution unique up to space-time translations,
starting simply from the physical soliton parameters $\{(a_n,c_n)\}_{n=1}^N$.

We begin with the following definition of the $N$-soliton parameter space.
\begin{definition}
An $N$-tuple of pairs $p_N:=\{(a_n,c_n)|a_n > 0\}_{n=1}^N \subset \Real^{2N}$,
of amplitudes and directions for the asymptotic line-solitons
associated with an $N$-soliton solution, is defined to be admissible if it 
yields the set $\Pi_N := \{(k_n^-,k_n^+)|\,k_n^\pm = \half(c_n \pm a_n)\}_{n=1}^N$ 
where the $2N$ phase parameters $\{k_n^{\pm}\}_{n=1}^N$ are distinct.
The set $Sol(N)$ of all admissible $N$-tuples of amplitudes and directions 
will be referred to as the $N$-soliton parameter space.
\label{D:parameter}
\end{definition}
Note that the pairs $(a_n,c_n),\, n=1,\ldots,N$ in the set $p_N$ are unordered. 
For example, $p_2=\{(1,2), (\half, 1)\}$ and $p_2= \{(\half, 1), (1,2)\}$ 
represent the same $2$-soliton solution. Similarly, $\Pi_N$ also consists
of unordered pairs $(k_n^-,k_n^+),\, n=1,\ldots,N$. However, since the parameters 
$\{k_n^{\pm}\}_{n=1}^N$ are distinct, they can be sorted in increasing order 
into an ordered set 
$K_{2N} = \{k_1, k_2, \ldots, k_{2N}\}$. Hence, $\Pi_N$ forms a partition of 
$K_{2N}$ into $N$ distinct pairs. The positions of each pair 
$(k_n^-,k_n^+) \in \Pi_N$ can be uniquely identified within $K_{2N}$ by an ordered 
pair of indices $[i_n,j_n]$ such that $i_n < j_n$. That is, 
$k_n^- = k_{i_n}$ and $k_n^+ = k_{j_n}$. It is precisely this identification 
that induces a correspondence between each $\Pi_N$ and a pairing map 
$\pi \in \mathcal{I}_{2N} \subset \mathcal{S}_{2N}$, the latter representing a disjoint partition 
of $[2N]$ into $N$ distinct pairs (see Property~\ref{p:Nsoliton}(iv)). It is
then clear from Definition~\ref{D:parameter} that this correspondence also
extends between each $p_N \in Sol(N)$ and a $\pi \in \mathcal{I}_{2N}$. 
\begin{example}
Consider a $3$-soliton parameter set
$p_3 = \{(a_1,c_1)=(1,-3), (a_2,c_2)=(\frac32, \half), (a_3,c_3)=(\frac32, \frac52)\}$,
and construct the set $\Pi_3=\{\half(c_n \pm a_n)\}_{n=1}^3$.
The set $p_3$ is admissible because the corresponding set 
$\Pi_3=\{(-2,-1),(-\half,1),(\half,2)\}$ contains six distinct
phase parameters. Sorting these parameters in increasing order yields
$K_6 = \{-2,-1,-\half,\half,1,2\} = \{k_1,k_2,k_3,k_4,k_5,k_6\}$.
Then, $\Pi_3=\{(k_1,k_2),(k_4,k_6),(k_3,k_5)\}$ which gives the
correspondence $p_3 \simeq \Pi_3 \mapsto (12)(35)(46) = \pi$.
\label{E:pi}
\end{example}
\noindent Note however that the correspondence between the soliton 
parameter space $Sol(N)$ and $I_{2N}$ is not one-to-one because distinct 
elements of $Sol(N)$ associated with solutions in the same equivalence class 
give rise to identical pairing. In this situation, the corresponding
sets $\Pi_{N}$ are distinct but after sorting, their elements are ordered
in identical fashion into the respective sets $K_{2N}$.
Thus, the soliton parameter
space~$Sol(N)$ is partitioned into disjoint sectors. Each sector 
corresponds to an equivalence class of solutions, distinguished
by an element $\pi \in \mathcal{I}_{2N}$, equivalently, by the set
$\{[e_n,g_n]\}_{n=1}^N $ labeling the $N$ asymptotic line-solitons.
The total number of such disjoint sectors
of $Sol(N)$ equals the cardinality $|\mathcal{I}_{2N}| = (2N-1)!!$.

Once a pairing map $\pi \in \mathcal{I}_{2N}$ is derived from a given $N$-soliton 
parameter set $p_N$, it is then possible to construct a coefficient matrix~$A$ 
satisfying Condition~\ref{positive}. Clearly, the pivot columns of $A$ 
will be labeled
by the excedance set $\{e_1,\ldots,e_N\}$ of $\pi$, and the non-pivot columns 
are labeled by $\{g_1,\ldots,g_N\}$, where $\pi(e_n) = g_n,\, n=1,\ldots,N$.
The explicit form of $A$ will be determined by using the rank conditions in 
Proposition~\ref{P:pairing}. 
Recall that in Examples \ref{E:2to1} and \ref{E:2to2}, we demonstrated how to
apply the results of Proposition~\ref{P:pairing} to a given coefficient matrix $A$,
and obtain the set of index pairs identifying the asymptotic
line-solitons. In the examples below, we will illustrate the reverse construction.
In other words, we will show that the rank conditions of Proposition~\ref{P:pairing}
are also {\em sufficient} to construct a coefficient matrix $A$ from a given
pairing map $\pi$ associated with the $N$-soliton solutions.
\begin{example}
We outline the construction of a coefficient matrix $A$ associated
with the $3$-soliton pairings $\{[1,2],[3,5],[4,6]\}$ found in 
Example~\ref{E:pi}. The construction proceeds in several steps.

\noindent\textit{Step~1.}~
It follows from Property~\ref{p:Nsoliton}(ii) that
the pivot and  non-pivot columns of the $3\times6$ matrix in RREF are labeled by
$\{e_1,e_2,e_3\} = \{1,3,4\}$, and $\{g_1,g_2,g_3\} = \{2,5,6\}$, respectively. 
So, the general form of $A$ satisfying Condition~\ref{positive} is
\begin{equation*}
A= \begin{pmatrix} 
 1 &z &0 &0 &v_1 &w_1 \\
 0 &0 &1 &0 &\!-v_2 &\!-w_2 \\ 
 0 &0 &0 &1 &v_3 &w_3 \\
\end{pmatrix}\,,
\end{equation*}
where $z, \{v_i, w_i\}_{i=1}^3$ are non-negative numbers to be determined.
Note that each unknown entry can be expressed as
certain maximal minors of $A$, (e.g., $A(234) =z$, etc).
The negative signs in the second row are included so that
the maximal minors of $A$ satisfy the non-negativity Condition~\ref{positive}(i).

\noindent\textit{Step 2.}~
In order to obtain further information about $A$, one needs to
apply the rank conditions in Proposition~\ref{P:pairing} to
the sub-matrices $X[ij]$ and $Y[ij]$ associated with each line
soliton $[i,j]$. 

If we start with the line-soliton $[1,2]$, and consider
the sub matrix $Y[12]=\emptyset$, we find that $\rank(Y[12])=0$.
Then according to Proposition~\ref{P:pairing}(ii), $\rank(Y[12]|1,2)=1$, 
so that columns $1$ and $2$ are proportional. Hence, $z\neq 0$.
The sub matrix $X[12]$ consists of columns $3,\ldots,6$
of $A$ above. From Proposition~\ref{P:pairing}(i), it follows that
$\rank(X[12]) \leq N-1=2$, since $N=3$. Then any maximal minor of $A$ consisting
of the columns of $X[12]$ will vanish. In particular, we find that
$A(345)=v_1 = 0$ and $A(346)=w_1 = 0$, so that
\begin{equation*}
A= \begin{pmatrix}
 1 &z &0 &0 &0 &0 \\
 0 &0 &1 &0 &\!-v_2 &\!-w_2 \\
 0 &0 &0 &1 &v_3 &w_3 \\
\end{pmatrix}\,,
\end{equation*}

\noindent\textit{Step 3.}~
Next we consider the index pair $[3,5]$. As the sub matrix $Y[35]=(0,0,1)^T$
is of rank 1, we obtain $\rank(Y[35]|5)=2$ from Proposition~\ref{P:pairing}(ii).
Therefore, columns 4 and 5 of $A$ are linearly independent. Then $v_2 \neq 0$.
The sub-matrix $X[35]$ consisting of the columns 1,2 and 6 of $A$, is of
rank 2, and its columns space is spanned by the linearly independent columns
$(1,0,0)^T$ and $(0,-w_2,w_3)^T$. Then from Proposition~\ref{P:pairing}(ii),
$\rank(X[35]|3)=3 \Rightarrow A(136)=w_3 \neq 0$, and 
$\rank(X[35]|5)=3 \Rightarrow A(156)=v_3w_2-v_2w_3 \neq 0$.
Moreover, the non-negativity of all maximal minors of $A$ requires that
$z, v_2, w_3$ and $A(156)$ must be positive. In particular, $A(156)=v_3w_2-v_2w_3 > 0$
implies that $v_3 \neq 0,\, w_2 \neq 0$, so these are also positive.
Thus, the matrix $A$ is parametrized by 5 positive parameters which can
be chosen as follows:\, $t_1=z, t_2=v_2, t_3=v_3, t_4=w_3, t_5= A(156)=v_3w_2-v_2w_3$.
It can be directly verified that all non-zero maximal minors of $A$ are polynomials 
in $t_1, \ldots, t_5$ with positive coefficients.

Thus, we have constructed a $5$-parameter family of coefficient matrix $A$ that
will generate the asymptotic line-solitons $[1,2],[3,5],[4,6]$ for any choice
of the positive parameter values $t_1, \ldots, t_5$, via Proposition~\ref{P:pairing}.
Note that the vanishing minors of $A$ satisfy the duality conditions of 
Eq.~\eqref{e:dualminor}.
Furthermore, any such $A$ together with the phase parameters in the set $K_6$ of 
Example~\ref{E:pi} would generate a $3$-soliton solution $u(x,y,t)$ with 
soliton parameters $p_3$. This solution is unique
up to space-time translations corresponding to different choices for the
phase constants $\theta_{m,0},\, m=1,\ldots 6$ in Eq.~\eqref{e:theta}.
\label{E:A3sol}
\end{example}
We make a few observations from the above example. First, note
that the rank conditions together with non-negativity of the maximal minors
completely determine which maximal minors of $A$ are zero, and which are non-zero.
Secondly, the non-vanishing maximal minors are chosen to be positive by expressing
them in terms of a suitable set of freely prescribed parameters which include
certain matrix elements of $A$ as well as certain combinations of the elements
of $A$. This parametrization completely determines the coefficient matrix $A$ 
which is in RREF and all the entries in its non-pivot columns can be
expressed as appropriate maximal minors. Thirdly, both rank
conditions (Propositions~\ref{P:pairing}(i) and (ii)) are applied
to the same index pair $[e_n,g_n]$ since it labels a pair of asymptotic 
line-solitons as $|y| \to \infty$. Instead, one could also apply the
rank conditions from either Proposition~\ref{P:pairing}(i) or (ii) to each
index pair, and recover the remaining information from the duality condition
Eq.~\eqref{e:dualminor}. For example, since columns 1 and 2 are proportional
for the coefficient matrix $A$ of Example \ref{E:A3sol}, the minors 
$A(125)=A(126)=0$. Then the duality condition implies that $A(346)=A(345)=0$
as we found in Step 2 above. We also point out here that in Ref.~\cite{Kodama},
the $N$-soliton $\tau$-function was required to satisfy a set of conditions,
which was referred to as the ``$N$-soliton condition'' (Definition 4.2 of ~\cite{Kodama}).
These conditions follow directly from the rank and duality conditions discussed in this 
article. We reiterate the above observations through another example.
\begin{example}
In this example, we construct a $4$-soliton solution starting with 
the following set of soliton parameters: 
$p_4 = \{a_n,c_n\}_{n=1}^4=
\{(1,-3), (\frac32,-\frac32), (\frac32,\half), (1,2)\} \in Sol(4)$.

\noindent {\em Step 1.}\, First, we construct the set 
$\Pi_4=\{(-2,-1),(-\frac32,0),(-\half,1) (\half,\frac32)\}$ from $p_4$
using $k_n^\pm = \half(c_n \pm a_n)$. Then we obtain the ordered set
$K_8=\{-2,-\frac32,-1,-\half,0,\half,1,\frac32\} = \{k_1,\ldots,k_8\}$
by sorting the phase parameters in $\Pi_4$ which can now be
re-expressed as $\Pi_4 = \{(k_1,k_3),(k_2,k_5),(k_4,k_7),(k_6,k_8)\}$.
This gives the correspondence $\Pi_4 \mapsto (13)(25)(47)(68) = \pi \in I_8$
where, $\pi$ is expressed as products of disjoint $2$-cycles. The asymptotic
line-solitons are identified by the set $\{[1,3],[2,5],[4,7],[6,8]\}$ of
index pairs.

\noindent {\em Step 2.}\, We proceed to construct the coefficient 
matrix $A$ that will generate the $\tau$-function of the $4$-soliton solution.
As in the previous example, we start with the $4 \times 8$ matrix $A$ 
satisfying Condition~\ref{positive}:
\begin{equation*}
A= \begin{pmatrix} 
 1 &0 &\!-z_1 &0 &u_1 & 0 &\!-v_1&\!-w_1 \\
 0 &1 &z_2 &0 &\!-u_2 & 0 &v_2& w_2 \\ 
 0 &0 &0 &1 &u_3 &0 &\!-v_3 &\!-w_3 \\
 0 &0 &0 &0 &0 &1 &v_4 &w_4   
\end{pmatrix}\,,
\end{equation*}
whose pivots and non-pivot indices are $\{e_1,e_2,e_3,e_4\} = \{1,2,4,6\}$, 
and $\{g_1,g_2,g_3,g_4\} = \{3,5,7,8\}$, respectively, and where 
$z_1,z_2,u_1,u_2,u_3$ and $v_i,w_i,\, i=1,\ldots,4$ are non-negative
reals to be determined. 

\noindent {\em Step 3.}\, We then 
apply the rank conditions from Proposition~\ref{P:pairing} 
systematically to each soliton index pair $[i,j]$, and collect all the
information regarding the unknown entries in terms of zero and non-zero 
minors of $A$. Here we indicate only the essential steps.

Consider the index pair $[1,3]$ and the associated sub-matrix $X[13]$ which 
consists of columns $4,\ldots,8$ of $A$ above. Since $\rank(X[13]) \leq N-1 = 3$, 
the maximal minors: $A(l_1,\ldots,l_4) =0, \, \{l_1,\ldots,l_4\} \subset \{4,5,6,7,8\}$.
For the index pair $[6,8]$, the sub-matrix $Y[68]=(-v_1,v_2,-v_3,v_4)^T$ is clearly
of rank 1. Then from Proposition~\ref{P:pairing}(ii) $\rank(Y[68]|6,8)=2$, which implies 
that $A(l,6,7,8)=0$ for any $l \in [8]$. Putting these conditions together, it can
be shown that all vanishing minors of $A$ obtained via
Proposition~\ref{P:pairing} are generated by the following relations
\begin{equation}
 u_iv_j-u_jv_i = 0, \qquad u_iw_j-u_jw_i = 0, \,\, i,j = 1,2, \qquad 
v_iw_j-v_jw_i = 0, \,\, i,j = 1,2,3 \,. 
\label{e:zerominors}
\end{equation}
Next we determine the non-zero entries of $A$. Note that $z_1 \neq 0$
since columns 2 and 3 of $A$ are linearly independent. This is because
the sub-matrix $Y[13] = (0,1,0,0)^T$ associated with index pair $[1,3]$, is of 
rank 1, then Proposition~\ref{P:pairing}(ii) implies that $\rank(Y[13]|3)=2$.
Consider now the sub-matrix $Y[47]$ whose rank is 2 since it is spanned by the 
linearly independent columns 5 and 6 of $A$. Then columns 4,5 and 6 are also
linearly independent as $\rank(Y[47]|4)=3$. Moreover, these columns also
span the column space of the sub-matrix $X[13]$ and form a basis which we
denote by $\beta_{13}$. By similar reasoning, we find the basis sets 
$\beta_{25}$ and $\beta_{47}$ for the column spaces of $X[25]$ and $X[47]$,
respectively. These are given by
\begin{equation*}
\beta_{13} = \begin{pmatrix}
 0 &u_1 & 0 \\
 0 &\!-u_2 & 0 \\
 1 &u_3 &0  \\
 0 &0 &1 
\end{pmatrix}\,, \qquad
\beta_{25} = \begin{pmatrix}
 1 &0 &-\!v_1 \\
 0 &0 &v_2 \\
 0 &0 &-\!u_3 \\
 0 &1 &v_4 
\end{pmatrix}\,, \qquad
\beta_{47} = \begin{pmatrix}
 1 &0 &-\!w_1 \\
 0 &1 &w_2 \\
 0 &0 &-\!w_3 \\
 0 &0 &w_4
\end{pmatrix}\,. \qquad
\end{equation*}
\begin{figure}[t!]
\centering
\raisebox{0.87in}{(a)}\includegraphics[scale=0.6]{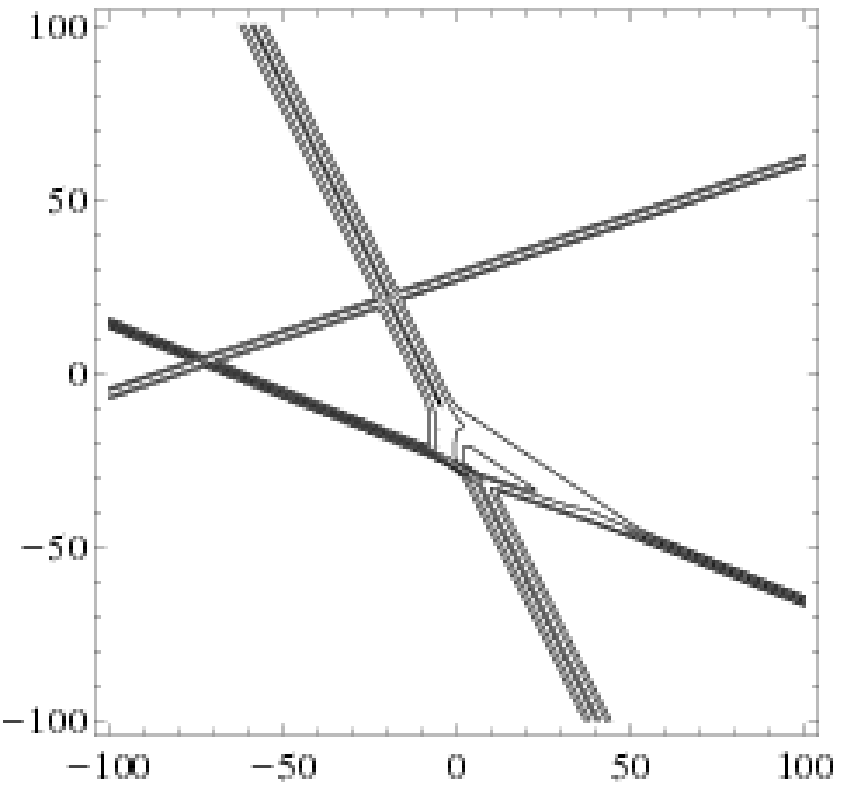} \hskip 0.6 in
\raisebox{0.87in}{(b)}\includegraphics[scale=0.6]{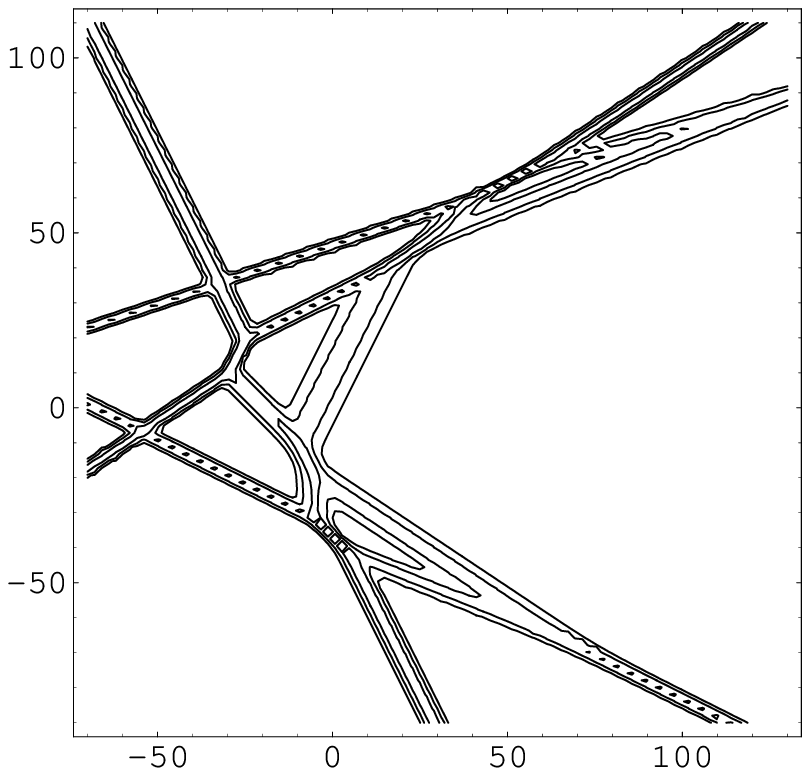} 
\caption{(a)~a $3$-soliton solution of Example~\ref{E:A3sol} at $t=12$, generated by the
coefficient matrix $A$ with parameter values: $z=v_2=w_2=1, \, v_3=3, \,w_3=\frac53$;
(b)~ a $4$-soliton solution of Example~\ref{E:A4sol} at $t=20$. The parameters of the
coefficient matrix $A$ are given by 
$z_1=u_1=u_2=v_1=v_2=v_3=1, \, z_2=u_3=v_4=w_1=w_2=w_3=2, \, w_4=3$. }
\label{f:Asol}
\end{figure}
Since both $\rank(X[13]|1)=\rank(X[13]|3)=4$, using the basis $\beta_{13}$
we get $A(1456)=u_2 > 0$, and $A(3456)=z_2u_1-z_1u_2 > 0$ where we also required
that the non-zero maximal minors to be positive. In addition, if we take $z_1 > 0$ 
(since $z_1 \neq 0$ from above), then $z_2>0,\, u_1 > 0$, as well. Proceeding in a 
similar fashion with $\beta_{25}$ and $\beta_{47}$ we find that that 
$u_3, v_2, v_3, v_4, w_3, w_4$ are all positive. Then from Eq.~\eqref{e:zerominors},
we finally conclude that all entries in the non-pivot columns of $A$ are non-zero
unlike the previous example. However, only 10 of the 13 non-zero elements of $A$
are independent due to the constraints from Eq.~\eqref{e:zerominors}.
Furthermore, by direct computation using the matrix elements of $A$ we find
that all vanishing minors of $A$ are in fact generated by the relations in
Eq.~\eqref{e:zerominors} which arise from the rank conditions 
of Proposition~\ref{P:pairing}.

As in Example~\ref{E:A3sol}, it is possible to construct a set of positive parameters
such that all non-zero maximal minors of $A$ are polynomials in these parameters with 
positive coefficients. These are given by
\begin{gather*}
t_1 = z_1, \qquad t_2=z_2, \qquad t_3 = \frac{u_1}{z_1}-\frac{u_2}{z_2}, \qquad
t_4 = \frac{u_2}{z_2}, \qquad t_5 = u_3, \\
t_6=\frac{v_2}{u_2}-\frac{v_3}{u_3}, \qquad t_7=\frac{v_3}{u_3}, \qquad 
t_8 = v_4, \qquad t_9 = \frac{w_4}{v_4}-\frac{w_3}{u_3}, \qquad t_{10}= \frac{w_4}{v_4} \,.  
\end{gather*}
\label{E:A4sol}
\end{example}
\noindent Figure~\ref{f:Asol}(a) shows a $3$-soliton solution with soliton parameters given 
in Example~\ref{E:pi} and a coefficient matrix $A$ constructed in
Example~\ref{E:A3sol}. Figure~\ref{f:Asol}(b) shows a $4$-soliton solution
with soliton parameters and matrix $A$ from Example~\ref{E:A4sol}.
It should be noted that the above examples are only illustrations 
of the general result asserting that for every pairing map $\pi \in \mathcal{I}_{2N}$ there
exists a parametrized family of $N \times 2N$ matrices $A$ which satisfy
Condition~\ref{positive} and generate an equivalence class of $N$-soliton solutions.
The proof of this remarkable result will be given in a future work~\cite{CK}.

\subsection{Combinatorics of $N$-soliton solutions}
Further refinement of the $N$-soliton classification scheme can be
achieved by studying the combinatorial properties of the associated 
$N \times 2N$ coefficient matrix $A$. We have already shown that the $N$-soliton 
solution space is characterized by the set $\mathcal{I}_{2N}$ of fixed point free involutions
of the permutation group $\mathcal{S}_{2N}$. In turn, these involutions of $\mathcal{S}_{2N}$ can
be enumerated in terms of the various possible arrangements of the pivot and
non-pivot columns of the $N$-soliton coefficient matrix $A$.  
A more geometric classification of the $N$-soliton solutions using
Schubert decomposition of the real Grassmannian $Gr(N,2N)$ has been carried
out in Ref.~\cite{Kodama} where the arrangements of pivot and non-pivot indices
were described in terms of Young diagrams. We resort to a more elementary
treatment here and present the main results below.
\begin{proposition}
Suppose that the index sets $\{e_1,\ldots,e_N\}$ and $\{g_1,\ldots,g_N\}$ with
$e_n < g_n,\,n=1,\ldots N$, form 
a disjoint partition of $[2N]$, and label respectively, the pivot and non-pivot
columns of a coefficient matrix $A$ associated with a $N$-soliton solution of KPII.
Then, the following results hold. \\
(i)\,\, The index set $\{e_1,\ldots,e_N\}$ is ordered as follows:
$1=e_1 < e_2 < \ldots < e_N < 2N$. Moreover, the elements satisfy
$n \leq e_n \leq 2n-1\,, \quad n=1,\ldots, N$. \\
(ii)\,\, The total number of choices $C_N$ for the ordered set $\{e_1,\ldots,e_N\}$ 
in item (i) is given by the $N$-th Catalan number (see e.g., \cite{Stanley}),
\begin{equation}\label{e:catalan}
 C_N = \frac{(2N)!}{N!(N+1)!}
\end{equation}
(iii)\,\, The set $\{g_1,\ldots,g_N\}$ is unordered, and the element $g_n$ can be
chosen in $2n-e_n$ ways for $n=1,\ldots,N$. Therefore, the number of possible choices for the unordered set
$\{g_1,\ldots,g_N\}$ for each set $\{e_1,\ldots,e_N\}$ is given by
\begin{equation}\label{e:numberg}
m(e_1,\ldots,e_N)=\prod_{n=1}^N(2n-e_n)\,.
\end{equation}
\label{P:combinatorics}
\end{proposition}
It follows from Proposition~\ref{P:combinatorics} that the total number
of distinct equivalence classes of $N$-soliton solutions satisfies the
curious combinatorial identity
\begin{equation}\label{e:FN}
F_N:= \sum_{\genfrac{}{}{0pt}{1}{e_1<\cdots<e_N,}{n \leq e_n \leq 2n-1}} 
m(e_1,\ldots,e_N)=
(2N-1)!! \,,
\end{equation}
which can be proved by following a similar line of argument that is provided
below in the proof of Proposition~\ref{P:combinatorics}(ii). Items (i) and (iii)
of Proposition~\ref{P:combinatorics} were already proved in Ref.~\cite{Kodama}.
\begin{proof}
({\em Proposition \ref{P:combinatorics}(ii)})\,
Let $E_N$ denote the set of all $N$-tuples $(e_1,\ldots,e_N)$ for which
Proposition \ref{P:combinatorics}(i) holds. Then it is clear that $|E_N|=C_N$.
Since $e_1=1$, each $N$-tuple contains one or more indices satisfying
$e_m=2m-1,\, m = 1,\ldots,N$. Then by sorting the elements of $E_N$ 
according to the {\em largest} positive integer $n \in [N]$ such that 
$e_n = 2n-1$, we obtain the disjoint partition 
$$ E_N = \bigsqcup_{n=1}^N W_n \qquad \mbox{where} \qquad
W_n = \left\{(e_1,\ldots,e_N)\,|\,e_n=2n-1, \quad e_m < 2m-1, \,\, m>n \right\} \,.$$
Note that $W_n$ can be expressed as the direct product:\,
$W_n = E_{n-1} \times \{e_n=2n-1\} \times \widehat{E}_{N-n}$, where
the set $E_{n-1} = \{(e_1,\ldots,e_{n-1})\}$ with $j \leq e_j \leq 2j-1,\,\, j \in [n-1]$, is 
defined similarly as $E_N$, and where
$\widehat{E}_{N-n} = \{(e_{n+1},\ldots,e_N)\}$ with 
$2n+j-1 \leq e_{n+j} < 2(n+j)-1, \,\, j \in [N-n]$. If we define new indices
$\widehat{e}_j := e_{n+j} - (2n-1),\,\, j=1,\ldots, N-n$ and relabel the elements of
$\widehat{E}_{N-n}$, then it should be clear that 
$$\widehat{E}_{N-n} \simeq 
\left\{(\widehat{e}_1,\ldots,\widehat{e}_j)\,|\quad j \leq \widehat{e}_j \leq 2j-1\right\}
=: E_{N-n} \,.$$
Now the cardinalities of $E_{n-1}$ and $\widehat{E}_{N-n}$ are $C_{n-1}$ and
$C_{N-n}$, respectively. Then it follows from above that $|W_n| = C_{n-1}C_{N-n}$, and
\begin{equation}
|E_n| = C_N = \sum_{n=1}^N C_{n-1}C_{N-n}
\label{e:CN}
\end{equation}
Eq.~\eqref{e:CN} gives a recursion relation for $C_N,\, N\geq 1$ with $C_0 := 1$. 
If $C(z) = \displaystyle\sum_{N=0}^{\infty}C_N\,z^N$ is the generating function
of the $C_N$, then Eq.~\eqref{e:CN} implies that $C(z)$ satisfies 
$z\,C^2(z) - C(z)+1 = 0$, yielding
$$ C(z) \,\, = \,\, \frac{1-\sqrt{1-4z}}{2z} \,\,  = \,\, 
\sum_{N=0}^{\infty}\frac{(2N)!}{N!(N+1)!}\,z^N \,,$$
by choosing the root consistent with $C(0) = C_0 =1$, then expanding it in power series.
Finally, we obtain the desired result by equating the coefficients of the two power series 
for $C(z)$.
\end{proof}

\begin{table}[b]
\begin{center}
\begin{tabular}{|c|c|c|}  \hline
$\|w\|$ &$w$ &$\{e_1,e_2,e_3\}$ \\
\hline
0 &$(0,0,0)$ &$\{1,2,3\}$ \\
\hline
1& $(0,0,1)$ & $\{1,2,4\}$ \\
\hline
2 &\begin{tabular}{c}
$(0,0,2)$ \\ $(0,1,1)$
\end{tabular}
& \begin{tabular}{c}
$\{1,2,5\}$ \\ $\{1,3,4\}$
\end{tabular}
\\ \hline
3 &$(0,1,2)$ &$\{1,3,5\}$ \\
\hline
\end{tabular}
\end{center}
\kern-\medskipamount
\caption{Weight vectors and possible pivot arrangements for 3-soliton solutions}
\label{t:pivot}
\end{table}
In view of Proposition~\ref{P:combinatorics}(i), it is natural to associate 
with each ordered set $\{e_1,\ldots,e_N\}$ a {\em weight vector} $w$
and its length $\|w\|$ defined by
\begin{equation}
w = (w_1,w_2,\ldots,w_N), \quad \mbox{where} \quad
w_n:= e_n-n \geq 0, \quad n \in [N], \quad \mbox{and} \quad
\|w\| = \sum_{n=1}^N(e_n-n)\,, 
\label{e:weight}
\end{equation}
respectively. Note that the weights form a non-decreasing sequence:\,
$0=w_1 \leq w_2 \leq \ldots \leq w_N$ and $0 \leq w_n \leq n-1$.
Similarly, we associate the unordered set of non-pivot indices
$\{g_1,\ldots,g_N\}$ with an {\em inversion} vector $\sigma$ defined by
\begin{equation}
\sigma = (\sigma_1, \sigma_2 \ldots \sigma_N), \quad \mbox{where} \quad
\sigma_n = |\{g_j\,|\,g_j > g_n,\,\, j<n\}|, \quad n \in [N]\,.
\label{e:inv}
\end{equation}
The inversions satisfy $0 \leq \sigma_n \leq 2n-e_n-1,\,\, n=1,\ldots,N$.
The upper limit of $\sigma_n$ follows from Proposition~\ref{P:combinatorics}(iii), 
by placing $g_n$ to the leftmost of the $2n-e_n$ available positions, and filling
the remaining positions with indices $g_j$ such that $j<n$. Notice that
the pair of vectors $(w,\sigma)$ are identical to the pair $(Y^+,Y^-)$ of
Young diagrams introduced in Ref~\cite{Kodama}.

\begin{figure}[h]
\centering
\includegraphics[scale=0.55]{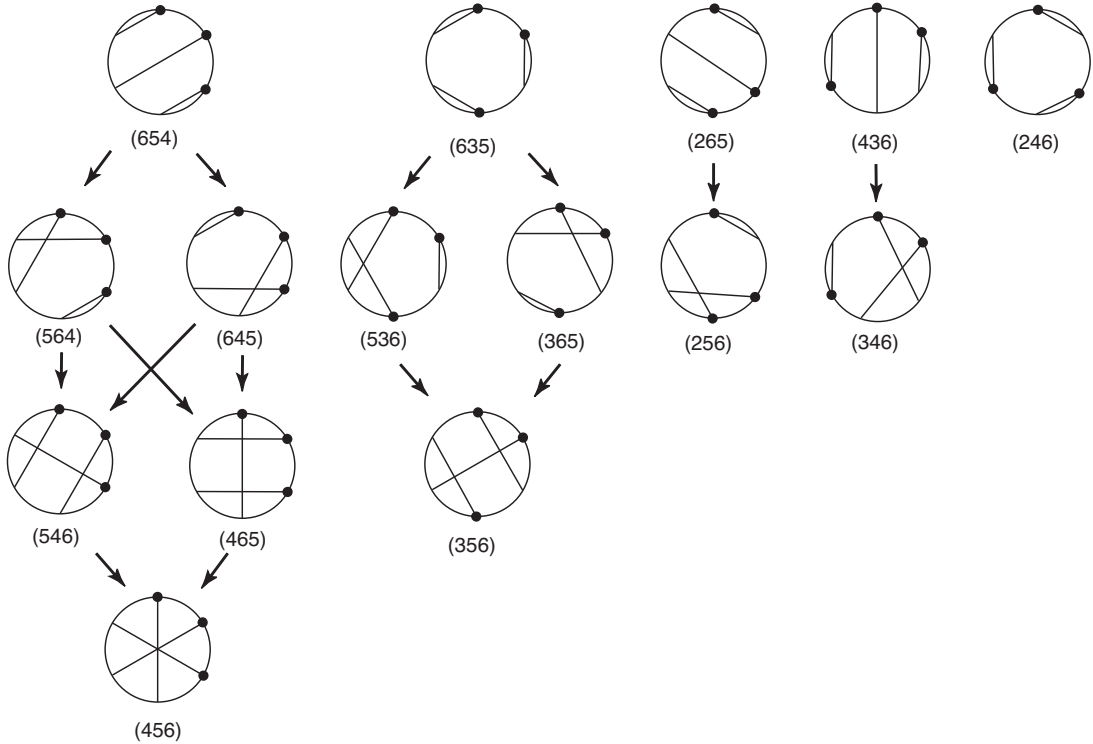}
\caption{The closed chord diagrams for $3$-soliton solutions. The dots on 
each diagram indicates the pivots $(e_1,e_2,e_3)$,
and the ordered numbers below the diagrams indicate $(g_1,g_2,g_3)$.
The number of the diagrams having the same number of crossings is
given by the generating function $F_3(q)=q^3+3q^2+6q +5$ where
5 is the Catalan number $C_3=F_3(0)$.}
\label{f:3chords}
\end{figure}

The results of Proposition~\ref{P:combinatorics} together with the weight and 
inversion vectors provide a refinement of the classification scheme for the 
$N$-soliton solutions. We illustrate here the refined scheme for $N=3$.
\begin{table}[h]
\begin{center}
\begin{tabular}{|c|c|c|c|} \hline
$\{e_1,e_2,e_3\}$ & $\{g_1,g_2,g_3\}$ & $\sigma$  & 3-soliton solution\\  
\hline
$\{1,2,3\}$  & \begin{tabular}{c}
             $\{4,5,6\}$ \\ $\{4,6,5\}$ \\ $\{5,4,6\}$ \\ 
             $\{5,6,4\}$ \\$\{6,4,5\}$ \\ $\{6,5,4\}$
               \end{tabular}
             &  \begin{tabular}{c}
             $(0,0,0)$ \\ $(0,0,1)$ \\ $(0,1,0)$ \\
             $(0,0,2)$ \\ $(0,1,1)$ \\ $(0,1,2)$
                 \end{tabular}
             &  \begin{tabular}{c}
             $\{[1,4],[2,5],[3,6]\}$ \\ $\{[1,4],[2,6],[3,5]\}$ \\ $\{[1,5],[2,4],[3,6]\}$ \\
             $\{[1,5],[2,6],[3,4]\}$ \\ $\{[1,6],[2,4],[3,5]\}$ \\ $\{[1,6],[2,5],[3,4]\}$
                  \end{tabular}
\\ \hline
$\{1,2,4\}$  & \begin{tabular}{c}  
             $\{3,5,6\}$ \\ $\{3,6,5\}$ \\ $\{5,3,6\}$ \\ $\{6,3,5\}$
               \end{tabular}
             & \begin{tabular}{c}
             $(0,0,0)$ \\ $(0,0,1)$ \\ $(0,1,0)$ \\ $(0,1,1)$
               \end{tabular}
             & \begin{tabular}{c}
             $\{[1,3],[2,5],[4,6]\}$ \\ $\{[1,3],[2,6],[4,5]\}$ \\  
             $\{[1,5],[2,3],[4,6]\}$ \\ $\{[1,6],[2,3],[4,5]\}$
                \end{tabular}
\\ \hline
$\{1,2,5\}$  & \begin{tabular}{c}
             $\{3,4,6\}$ \\ $\{4,3,6\}$
               \end{tabular}
             & \begin{tabular}{c}
             $(0,0,0)$ \\ $(0,1,0)$
               \end{tabular}
             & \begin{tabular}{c}
             $\{[1,3],[2,4],[5,6]\}$ \\ $\{[1,4],[2,3],[5,6]\}$  
                \end{tabular}
\\ \hline
$\{1,3,4\}$  &\begin{tabular}{c}
             $\{2,5,6\}$ \\ $\{2,6,5\}$
               \end{tabular}
             & \begin{tabular}{c}
             $(0,0,0)$ \\ $(0,0,1)$
               \end{tabular}
             & \begin{tabular}{c}
             $\{[1,2],[3,5],[4,6]\}$ \\ $\{[1,2],[3,6],[4,5]\}$
                \end{tabular}
\\ \hline
$\{1,3,5\}$  & $\{2,4,6\}$  & $(0,0,0)$  & $\{[1,2],[3,4],[5,6]\}$ \\ \hline
\end{tabular}
\end{center}
\caption{ The 15 distinct 3-soliton solutions } 
\label{t:nonpivot} 
\end{table}
In this case, $A$ is a $3 \times 6$ matrix with 3 pivots
satisfying $e_1 =1, \,\, 2 \leq e_2 \leq 3, \,\, 3 \leq e_3 \leq 5$.
From Proposition~\ref{P:combinatorics}(ii), the total number of pivot
configurations $\{e_1,e_2,e_3\}$ is given by the Catalan number
$C_3 = 5$. Thus, there are 5 subclasses of $3$-soliton solutions depending
on $5$ distinct pivot configurations which are determined
by the associated weight vectors. These subclasses are listed in
Table~\ref{t:pivot} in increasing order of the length $\|w\|$ of the
weight vector.
For each of these pivot arrangements, Proposition~\ref{P:combinatorics}(iii)
gives the total number of distinct non-pivot configurations. Each set
$\{g_1,g_2,g_3\}$ in a given subclass, is distinguished by its unique inversion
vector. The various 3-soliton solutions in each subclass is shown in
Table ~\ref{t:nonpivot} and the associated chord diagrams are presented
in Figure~\ref{f:3chords}. Note that instead of open chord diagrams, we use
circles in Figure~\ref{f:3chords}. Here each straight chord replaces the
pair of upper and lower chords connecting the same index pair $[e_n,g_n]$ in the 
self-dual, open chord diagrams of the $N$-soliton solutions 
(cf. Property~\ref{p:Nsoliton}(iv)).

The subclass associated with the pivot configuration
$\{1,2,3\}$ is isomorphic to the permutation group $\mathcal{S}_3$ acting on 
$\{4,5,6\}$ to form the non-pivot sets $\{g_1,g_2,g_3\}$. These are
arranged in the second column of Table ~\ref{t:nonpivot} according to
the non-decreasing order of the $L^1$-norm $|{\bf \sigma}|=\sigma_1+\sigma_2+\sigma_3$ 
of the respective inversion vectors in the second column 
of Table ~\ref{t:nonpivot}. The corresponding set of diagrams forming a
hexagon in Figure~\ref{f:3chords} is the permutahedron for $\mathcal{S}_3$.
On the other hand when $\{e_1,e_2,e_3\} = \{1,2,4\}$ the non-pivot index $g_3$
is chosen in $2 \times 3-4=2$ ways; $g_2$ is chosen in $2 \times 2-2=2$ ways;
and obviously there is only one way to choose $g_1$. Thus, there are only
4 (instead of 6) possible ways the non-pivot columns 3,5 and 6 can be arranged to
form the set $\{g_1,g_2,g_3\}$. This is due to the restriction that $g_3 \neq 3$
because according to Proposition~\ref{P:combinatorics}, $g_3$ must be greater 
than $e_3=4$. The chord diagrams with the pivot set $\{1,2,4\}$ form
a square which is a subpolytope of the permutahedron of $\mathcal{S}_3$. We will discuss
the polytope structure for the $N$-soliton solutions in \cite{CK}.

With the circular chord diagrams for the sets $\{e_1,\ldots,e_N\}$ and $\{g_1,\ldots,g_N\}$,
one can easily find a $q$-analog of the function $m(e_1,\ldots,e_N)$ in Eq.~\eqref{e:numberg}
defined by
\[
m(e_1,\ldots,e_N)(q)=\prod_{n=1}^N[2n-e_n]_q=\sum_{c=0}^{c_{max}}m_cq^c\,.
\]
Here $m_c$ gives the number of the circular chord diagram having $c$ crossings, and
the maximum number of crossings for given $\{e_1,\ldots,e_N\}$ is~\cite{CK}
\[c_{max} = N^2-\sum_{n=1}^Ne_n \,.
\]
For example, when $\{e_1,e_2,e_3\}=\{1,2,3\}$ and $\{e_1,e_2,e_3\}=\{1,2,4\}$, 
we have 
\[
m(1,2,3)(q)=q^3+2q^2+2q+1 \qquad \mbox{and} \qquad  m(1,2,4)(q)=q^2+2q+1 \,,
\]
corresponding to the hexagon and square in Figure~\ref{f:3chords}.
Note that $m(e_1,\ldots,e_N)(q=1)=m(e_1,\ldots,e_N)$. 
One can also define a $q$-analog of the function $F_N$ in Eq.~\eqref{e:FN} which
gives the number of circular chord diagrams having $c$ crossings for given $N$, i.e.
the number of $N$-soliton solutions having $c$ T-type interactions among $N$ line solitons:
\[
F_N(q):=\sum_{\{e_1,\ldots,e_N\}}m(e_1,\ldots,e_N)(q)\,.
\]
For example, we have $F_3(q)=q^3+3q^2+6q+5$ as is evident from counting the number
of circular chord diagrams (from bottom to top) at each crossing level in 
Figure~\ref{f:3chords}.
Note also that $F_N(1)=F_N$ and $F_N(0)=C_N$, the $N$-th Catalan number which
equals the number of possible pivot configurations $[\{e_1,\ldots,e_N\}]$, as well as
the number of chord diagrams with no crossings. The latter case yields the number
of $N$-soliton solutions with only P- and O-type interactions among the $N$ line solitons.
Furthermore, if we define the generating function by
\[
F(q,x):=\sum_{N=0}^{\infty}F_N(q)\, x^N\,,
\]
then it is possible to show that $F(q,x)$ can be expressed by the continued fraction 
of the Stieltjes type,
\[F(q,x)=\cfrac{1}{1-
          \cfrac{x\,[1]_q}{1-
           \cfrac{x\,[2]_q}{1- 
            \cfrac{x\,[3]_q}{1-\cdots}}}} \,,
\]
with $F_0(q)=1$. A proof will be given in \cite{CK} (see also \cite{Kasraoui}).

\section*{Acknowledgments}

SC thanks Gino ~Biondini for useful discussions and some figures.
This work was partially supported by the National Science Foundation 
Grant No. DMS-0307181 and Grant No. DMS-0404931.

\catcode`\@ 11
\def\journal#1&#2,#3 (#4){\begingroup \let\journal=\d@mmyjournal {\frenchspacing\sl #1\/\unskip\,} {\bf\ignorespaces #2}\rm, #3 (#4)\endgroup}
\def\d@mmyjournal{\errmessage{Reference foul up: nested \journal macros}}
\def\title#1{{``#1''}}
\def\@biblabel#1{#1.}
\catcode`\@ 12

\end{document}